%% file: main.tex
\documentclass[a4paper]{scrartcl}
\usepackage{graphicx}
\usepackage{booktabs}
\usepackage{enumitem}

\usepackage{amsmath}
\usepackage{amsfonts,amssymb}
\usepackage[amsmath,thmmarks,standard]{ntheorem}

\usepackage{bm}
\usepackage{fixmath}

\usepackage{hyperref}
\hypersetup{
  colorlinks=false,
  frenchlinks=false,
  pdfborder={0 0 0},
  naturalnames=false,
  hypertexnames=false,
  breaklinks
}
\usepackage{xcolor}
\usepackage{graphics,graphicx}

\usepackage[capitalize,nameinlink]{cleveref}

\crefname{section}{section}{sections}
\crefname{subsection}{subsection}{subsections}
\Crefname{section}{Section}{Sections}
\Crefname{subsection}{Subsection}{Subsections}

\Crefname{figure}{Figure}{Figures}

\crefformat{equation}{\textup{#2(#1)#3}}
\crefrangeformat{equation}{\textup{#3(#1)#4--#5(#2)#6}}
\crefmultiformat{equation}{\textup{#2(#1)#3}}{ and \textup{#2(#1)#3}}
{, \textup{#2(#1)#3}}{, and \textup{#2(#1)#3}}
\crefrangemultiformat{equation}{\textup{#3(#1)#4--#5(#2)#6}}%
{ and \textup{#3(#1)#4--#5(#2)#6}}{, \textup{#3(#1)#4--#5(#2)#6}}{, and \textup{#3(#1)#4--#5(#2)#6}}

\Crefformat{equation}{#2Equation~\textup{(#1)}#3}
\Crefrangeformat{equation}{Equations~\textup{#3(#1)#4--#5(#2)#6}}
\Crefmultiformat{equation}{Equations~\textup{#2(#1)#3}}{ and \textup{#2(#1)#3}}
{, \textup{#2(#1)#3}}{, and \textup{#2(#1)#3}}
\Crefrangemultiformat{equation}{Equations~\textup{#3(#1)#4--#5(#2)#6}}%
{ and \textup{#3(#1)#4--#5(#2)#6}}{, \textup{#3(#1)#4--#5(#2)#6}}{, and \textup{#3(#1)#4--#5(#2)#6}}

\newtheorem{assumption}{Assumption}
\crefname{assumption}{Assumption}{Assumptions}

\newcommand{\pmx}[1]{\begin{pmatrix}#1\end{pmatrix}}
\newcommand{\smx}[1]{\left(\begin{smallmatrix}#1\end{smallmatrix}\right)}
\newcommand{\Real}{\mathbb R}
\newcommand{\ones}{\mathbf{1}}
\renewcommand{\phi}{\varphi}
\renewcommand{\v}[1]{\mathbold{#1}}
\newcommand{\vv}[1]{\mathbold{#1}}
\newcommand{\eye}{\mathbold{I}}
\DeclareMathOperator{\diag}{diag}
\DeclareMathOperator{\VEC}{Vec}
\DeclareMathOperator{\E}{E}
\DeclareMathOperator{\Prob}{P}
\DeclareMathOperator{\Cov}{Cov}

\DeclareMathOperator{\tr}{tr}
\title{Multifactor Quadratic Hobson and Rogers models%
}%
\author{%
  Paolo Foschi\thanks{%
    Dept. of Statistical Sciences, University of Bologna
    (\href{mailto:paolo.foschi2@unibo.it}{paolo.foschi2@unibo.it}).}
}
\date{}

\begin{document}

\maketitle

% ----------------------------------------------------------------
\begin{abstract}
  A multi-factor extension of the Hobson and Rogers (HR) model,
  incorporating a quadratic variance function (QHR model), is proposed
  and analysed. The QHR model allows for greater flexibility in
  defining the moving average filter while maintaining the Markovian
  property of the original HR model. The use of a quadratic variance
  function permits the characterisation of weak-stationarity
  conditions for the variance process and allows for explicit
  expressions for forward variance.
  Under the assumption of stationarity, both the variance and the
  squared increment processes exhibit an ARMA autocorrelation
  structure. The stationary distribution of the prototypical scalar
  QHR model is that of a translated and rescaled Pearson type IV
  random variable.
  A numerical exercise illustrates the qualitative properties of the
  QHR model, including the implied volatility surface and the term
  structures of forward variance, at-the-money (ATM) volatility, and
  ATM skew.
\end{abstract}

% \begin{keywords}

\textbf{MSC:}
% \begin{MSCcodes}
60H10, 60G10, 91G20, 91G30
% \end{MSCcodes}
  
% ----------------------------------------------------------------

%%
\section{Introduction}

In recent years there has been a growing interest in path dependent
volatility (PDV) continuous time models, primarily driven by the need
of pricing VIX derivatives
\cite{GatheralJusselinRosenbaum:2020,Guyon:2022}.
The Hobson and Rogers (HR) model was one of the first models in that
class \cite{HobsonRogersR:98}. In the HR model the volatility is a
deterministic function of the offset of the current log-price level
$x_t = \log(S_t)$ with respect to its exponential moving average. More
precisely, the HR offset is defined as
\begin{align*}
  y_t &= x_t - \int_{-\infty}^t \phi(t-s)x_s ds,
  &
  \phi(t) &= \lambda e^{-\lambda t},
\end{align*}
where $\lambda > 0$, and the risk-neutral dynamics of $x_t$ is
\begin{align}\label{eq:intro:dx}
  x_t = -\frac12 \sigma_t^2dt + \sigma_t dW_t
\end{align}
where $\sigma_t^2 = f(y_t)$ for some function $f: \Real \to \Real_+$
that links the offset to the underlying spot variance. One of the best
features of the HR approach is that the dynamics of $x_t$ and $y_t$ is
Markovian. Indeed, the offset process solves the stochastic
differential equation (SDE)
\begin{align}\label{eq:intro:dy}
  dy_t = -\lambda y_tdt + dx_t .
\end{align}
Still, that approach leaves some freedom to the practitioner by
allowing the choice of the memory parameter $\lambda$ and of the
function $f$ that links the offset level to the instantaneous
variance. The performances of that approach have been tested by Foschi
and Pascucci \cite{FoschiPascucci:2009}.

Another advantage in using the HR model is the fact that the model
dynamics, and in turn, the resulting pricing functions, are
``non-dimensional'' \cite{LiptonReghai:2023}. Indeed, the SDEs
\cref{eq:intro:dx} and \cref{eq:intro:dy} are autonomous, since their
coefficients do not depend on the time variable, and the link function
only depends on the offset $y_t$. Being the difference of a log-price
and an average of log-prices, $y_t$ is a non-dimensional quantity
(analogous to a log-ratio).

The downside of the HR models is the excessive rigidity in the choice
of moving average filter $\phi$. A tentative generalisation in that
direction is represented by the so called Path Dependent Volatility
model (PDV). In that model, a different filter function can be chosen,
but at the cost of obtaining SDEs that are not autonomous, i.e. their
coefficients depend on absolute time
\cite{FoschiPascucci:2008,HalulliVargiolu:2008}. As a consequence, PDV
model parameters tend to be heavily time- and contingency-dependent.

The purpose of this work is to generalise the HR model to a parametric
model which allows more freedom in the choice of the moving average
filter without loosing the advantages autonomous SDEs and a
non-dimensional volatility specification. A second modelling objective
is to obtain explicit expressions of the conditional expectations of
the first two moments of the variance process.

These results allow to obtain the autocorrelation structure of the
variance process allowing to characterise the weak-stationarity of the
spot variance process. With respect to those issues, notice that, the
principal component analysis, which is often used in finance to
present stylised facts and justify term structure models, requires
weak-stationarity \cite{ZhangTong:2022}.
Moreover, mean reversion and stationarity represent major selling
points in the marketing of variance related products like
VIX\footnote{%
  See CBOE's VIX page
  \url{https://www.cboe.com/tradable_products/vix/}. }.

Under the assumptions of weak-stationarity, the variance process and
the squared increments have the autocovariance structure of an ARMA
process \cite{Brockwell:2004}. But, differently from
\cite{BarndorffNielsen:2000} or \cite{BrockwellChadraaLindner:2006},
that structure is here obtained without introducing additional
Brownian motions and remaining in a diffusive setup. 

As an additional advantage, the model considered here allows for an
explicit formula of the conditional variance. In other words, the term
structure of forward variance can be explicitly computed without the
need of a Monte Carlo integration, simplifying model calibration
tasks.

\bigskip

Here, the HR offset $y_t$ is generalised to a $p$-dimensional process
having dynamics
\begin{align}\label{eq:intro:dvy}
  d\v{y}_t = -\vv\Lambda \v{y}_t dt + \v{b}\sigma_t dW_t
\end{align}
where $\v{b} \in \Real^p$, all the eigenvalues of
$\vv\Lambda \in \Real^{p\times p}$ have positive real part.
That process corresponds to exponential moving average:
\begin{align*}
  \v{y}_t =  \int_{-\infty}^t \vv\Lambda e^{-\vv\Lambda (t-s)} \v{b}(\xi_t - \xi_s) ds,
\end{align*}
where $\xi_t$ is the process with dynamics
$d\xi_t = \sigma_t dW_t = dS_t/S_t$. Note that, in \cref{eq:intro:dy}
the offset process was averaging the log-price innovations $dx_t$,
while in \cref{eq:intro:dvy} the average is performed only on their
unpredictable part $\sigma_tdW_t = dS_t/S_t$. Next the offsets in
$\v{y}_t$ can be linearly combined to obtain an offset process
$\tilde{y}_t = \v{w}^T\v{y}_t$, $\v{w} \in \Real^p_+$. When
$\v{b}^T\v{w} = 1$, that choice corresponds to an offset computed
w.r.t. a moving average with a filter that depends on $\vv\Lambda$,
$\v{b}$ and $\v{w}$:
\begin{align}\label{eq:intro:tildey}
  \tilde{y}_t &= \xi_t - \int_{-\infty}^t \phi(t-s) \xi_s ds,
  &
  \phi(t) &= \v{w}^T\vv\Lambda e^{-\vv\Lambda t}\v{b}.
\end{align}
Note that $\phi(t)$ is an exponential polynomial:
$\phi(t) = \sum_i t^{n_i}e^{\alpha_i -\beta_i t}$ for some finite set
of coefficients $\{\alpha_i, \beta_i \in \mathbb{C}\}$. 

The second element of the HR modelling consists in the specification
of the variance link function. In order to contain the number of
parameter and to obtain a analytically tractable model the spot variance is
defined as a quadratic function of $\v{y}_t$:
\begin{align}\label{eq:intro:vol0}
  \sigma_t^2 = \alpha + 2\v\beta^T\v{y}_t + \v{y}_t^T\vv\Gamma\v{y}_t,  
\end{align}
where $\alpha > 0$, $\vv\Gamma \in \Real^{p\times p}$ and
$\v\beta \in \Real^p$ is such that $\sigma_t^2 > 0$ for all
$\v{y}_t \in \Real^p$. It will be shown that, with this choice of the
variance link, the variance mean-reversion, its conditional
expectation and weak-stationarity can be explicitly characterised.

Again, when $\v\beta = \beta_0\v{w}$ and
$\vv\Gamma = \gamma_0\v{w}\v{w}^T$, the spot variance is a quadratic
function $\tilde{y}_t$. Thus, the model given by \cref{eq:intro:dvy}
and \cref{eq:intro:vol0}, encompass HR models where the moving
average filter is generalised to an exponential polynomial.

Models based on exponential polynomial filters have been considered in
\cite{AbiJaberElEuch:2019,Buehler:2006,GazzaniGuyon:2024,GuyonLekeufack:2023}.
In particular, multifactor diffusion models have been used to
approximate Rough Volatility models
\cite{GazzaniGuyon:2024,GuyonLekeufack:2023}. The Quadratic Rough
Heston model is analogous the one here proposed in using a quadratic
function to link a path moving average to the instantaneous variance
coefficient \cite{GatheralJusselinRosenbaum:2020}. For a review on
rough and non-rough volatility models see \cite{DiNunnoEtAl:2024}.
Note that the actual roughness of financial volatility processes is
not a settled question \cite{ContDas:2024}. Also, discrete time models
where the volatility depends on different filters (different scales),
have been around for some time. For a couple of different successful
approaches, see the Component GARCH and the Realised Volatility models
\cite{Christoffersen:2008,Corsi:2009}.

\bigskip

The paper is structured as follows. In \cref{sec:model}, the model is
presented with some results on its solution. Identification of the
model is briefly discussed in \cref{sec:ident}. Then, in
\cref{sec:homo}, homogeneity properties of the transition density and
pricing functions are discussed. Next, \cref{sec:filters}
characterises the model as a path-dependent volatility one where the
latent factor consists on auto-regressive (exponential) filters. Then,
in \cref{sec:conditionalMoments}, conditional moments of the offset
and variance processes are characterised in terms of matrix
exponentials. That characterisation allows to obtain sufficient
condition for the mean reversion or weak-stationarity of the variance
and conditional variance processes (\cref{sec:weakStationarity}). A
simple algorithm for computing the shape of the principal component
decomposition of the forward variance term structure is presented in
\cref{sec:fwdVariance}.

In \cref{sec:tests}, a couple of special cases are presented with an
extensive set of numerical examples showing the characteristics and
potential flexibility of this class of models. The first model
considered is the scalar (classical) QHR model. Given its simplicity,
that model allows for weak-stationarity conditions to be explicitly
defined. Moreover, it is shown that under those conditions there
exists a stationary solution for the offset-process $y_t$. Moreover,
the stationary distribution for $y_t$ consists on a Pearson type IV
distribution \cite{Pearson:1895}, that reduces to a scaled student-t
distribution for the symmetric case $\beta=0$. The second class of
models presented in Section 3 are the rank-one models. In this class
of models, the instantaneous variance is function of a single offset
process obtained as in \cref{eq:intro:tildey}. In that case, we could
not simplify further the result obtained in \cref{sec:model} and their
analysis only consists on a numerical one.

An additional section, \cref{sec:proofs}, contains the proofs
of the results reported in the previous sections.

\subsection{Notation}\label{sec:notation}

Most of the notational conventions used in the rest of this work are
here introduced. The set of $n$-dimensional real vectors and
$m\times n$ real matrices are denoted by $\Real^n$ and
$\Real^{m\times n}$, respectively. Vectors are represented using
lowercase bold symbols and matrices by means of bold capital
characters. Vectors are always meant to be column vectors, that is an
$n$-vector is a $n\times 1$ matrix. The transpose of a matrix $\vv{A}$
is denoted by $\vv{A}^T$. The identity matrix and its $i$-th column
are denoted by $\eye$ and $\v{e}_i$, respectively. Their dimensions
should be deduced from the context. If necessary the $n \times n$
identity will be denoted by $\eye_n$. The determinant and the trace
operators are denoted by $\tr(\vv{A})$ and $\det(\vv{A})$,
respectively. If not explicitly specified, vector norms are 2-norms:
$\|\v{x}\| = \|\v{x}\|_2 = (\v{x}^T\v{x})^{\frac12}$ for
$\v{x}\in \Real^n$, and matrix norms are induced 2-norms:
$\|\vv{A}\| = \max_{\v{x}} \|\vv{A}\v{x}\|_2/\|\v{x}\|_2$.
% The matrix Frobenious norm is defined as
% $\|\vv{A}\|_F = \| \VEC(\vv{A})\| = \sqrt{\tr(\vv{A}^T\vv{A})}$.
% Expectations and conditional expectations under $\Prob$ are denoted by
% $\E[\cdot]$ and $\E[\cdot \;|\; \cdot]$.
The expectation of a vector (matrix) consists on the vector (matrix)
of the expectations.

The sets of real $m \times n$-dimensional matrices with non-negative
elements is denoted by $\Real_+^{m \times n}$ and that of $n\times n$
real, symmetric and positive semi-definite matrices by
$\mathcal{S}^n_+$. Let $\v{A} \in \Real^{m \times n}$, $\v{A} \geq 0$
means that $\vv{A} \in \Real_+^{m \times n}$ (element-wise).
Analogously, $\v{A} \geq \v{B}$ if and only if $\v{A} - \v{B} \geq 0$.

If $\vv{A} \in \Real^{m\times n}$, $\VEC(\vv{A}) \in \Real^{mn}$ is
the vector obtained by stacking the columns of $\vv{A}$.
% The inverse of the operator $\VEC:\Real^{m \times n} \to
% \Real^{mn}$ is denoted by $\VEC^{-1}_{m\times n}$.
The Kronecker product of $\vv{A}\in \Real^{m\times n}$ and
$\vv{B} \in \Real^{p\times q}$ is represented by
$\vv{A} \otimes \vv{B}$ and has dimensions $mp \times nq$. Moreover,
$\vv{A}^{\otimes 0} = 1$ and
$\vv{A}^{\otimes k)} = \vv{A} \otimes \vv{A}^{\otimes (k-1)}$ for
$k\geq 1$ and $\vv{A}\in\Real^{m\times n}$. The definitions of the
$\VEC$ and Kronecker product operators and a review of their
properties can be found in \cite{MagnusNeudecker:book}.

\section{The QHR Model}
\label{sec:model}

The QHR model is specified as follows. The (risk-neutral) dynamics of
the discounted log-price $x_t$ of the latent vector $\v{y}_t$ is given
by
\begin{subequations}\label{eq:mv:dyn}
\begin{align}
  \label{eq:mv:dx}
  d x_t &= - \frac12 \sigma_t^2 dt + \sigma_t dW_t,
  \\
  \label{eq:mv:dy}
  d\v{y}_t &= -\vv\Lambda \v{y}_t dt + \v{b} \sigma_t dW_t,
  \\
  \label{eq:mv:vol0}
  \sigma_t^2 &= \alpha + 2\v{y}_t^T\v\beta + \v{y}_t^T\vv\Gamma \v{y}_t,
\end{align}
\end{subequations}
where $W_t$ is a standard Brownian motion on a filtration
$(\mathcal{F}_t)_{t\geq 0}$ of a probability space
$(\Omega,\mathcal{F},\Prob)$. The parameters
$\vv\Lambda \in \Real^{p\times p}$, $\v{b} \in \Real^p$,
$\alpha\in \Real$, $\v\beta \in \Real^p$ and
$\vv\Gamma \in \Real^{p\times p}$ satisfy the following assumption
which is sufficient to guarantee that the SDEs \cref{eq:mv:dx} and
\cref{eq:mv:dy} has a unique non-exploding solution in finite time
(see \cref{thm:mv:standardHyp} below).

\begin{assumption}\label{h:1}

  The eigenvalues of $\vv\Lambda$ are real and positive, $\alpha > 0$,
  $\smx{\alpha & \v\beta^T \\ \v\beta & \vv\Gamma}$ is positive
  semidefinite and $(x_0,\v{y}_0)$ is $\mathcal{F}$-measurable with
  $\E[ x_0^2 + \|\v{y}_0\|^4] < \infty$.
\end{assumption}

\begin{lemma}\label{thm:mv:standardHyp}
  Under \cref{h:1} the SDE \cref{eq:mv:dyn} has a unique non-exploding
  in finite time solution.

  \begin{proof}
    The proof consists on exploiting the standard hypothesis of a
    system of SDEs. While the coefficients of the system of SDEs
    \cref{eq:mv:dx} and \cref{eq:mv:dy} do not satisfy the standard
    hypothesis, adding to that system the dynamics of
    $\v{q}_t = \v{y}_t \otimes \v{y}_t$ results in a system of SDEs
    whose coefficients are Lipschitz continuous and linearly bounded.

    Indeed,
    \begin{align}
      \label{eq:mv:dq}
      d\v{q}_t &= (\bar{\v{b}}\sigma_t^2 - \bar{\vv\Lambda} \v{q}_t)dt
      + (\v{b}\otimes \v{y}_t + \v{y}_t \otimes \v{b})\sigma_t dW_t,
      &\text{and}&&
      \sigma_t^2 &= \alpha + 2\v{y}_t^T\v\beta + \v\gamma^T\v{q}_t
    \end{align}
    where $\bar{\v{b}} = \v{b}\otimes \v{b}$,
    $\bar{\vv\Lambda} = \vv\Lambda \otimes \eye_p + \eye_p \otimes
    \vv\Lambda$ and $\v\gamma = \VEC(\vv\Gamma)$. Thus, since
    $\sigma_t^2$ is an affine function of the state process
    $(x_t; \v{y}_t; \v{q}_t)$, the trend coefficients of the system
    \cref{eq:mv:dyn} and \cref{eq:mv:dq} are affine
    functions of the state processes. Moreover, the diffusion
    coefficients of $x_t$, $\v{y}_t$ and $\v{q}_t$ are, respectively,
    given by $\sigma_t$, $\v{b}\sigma_t$ and
    $(\v{b}\otimes \v{y}_t + \v{y}_t \otimes \v{b})\sigma_t$. Since
    $\alpha > 0$ by \cref{h:1}, the $\sigma_t$ and, in turn,
    $\v{b}\sigma_t$, are Lipschitz function of $\v{y}_t$. Regarding
    the diffusion coefficient of $\v{q}_t$, its squared norm is given
    by
    \begin{align*}
      \| (\v{b}\otimes\v{y}_t + \v{y}_t\otimes\v{b})\|^2\sigma_t^2
      &=
      2\Big(\v{b}^T\v{b} \, \v{y}_t^T\v{y}_t + (\v{b}^T\v{y}_t)^2\Big)
      \sigma_t^2
      \\
      &\leq (A_1 + A_2\| \v{q}_t\|)\sigma_t^2
      \leq (B_1 + B_2 \|\v{q}_t\|^2),
    \end{align*}
    for some positive constants $A_1,A_2,B_1$ and $B_2$. It follows
    that this diffusion coefficient is Lipschitz with at most linear
    growth. Consider also the assumption \cref{h:1}.c, the system of
    SDEs \cref{eq:mv:dyn} and \cref{eq:mv:dq} satisfies the standard
    hypothesis and thus it has a unique non-exploding in finite time
    solution. By construction, the system of SDEs \cref{eq:mv:dyn} and
    \cref{eq:mv:dq}, with initial condition
    $\v{q}_0=\v{y}_0\otimes \v{y}_0$, coincides with the system
    \cref{eq:mv:dyn}. It follows that also the latter has a unique
    non-exploding in finite time solution.
  \end{proof}
\end{lemma}

Here, the dynamics \cref{eq:mv:dyn} has not been deliberately placed
under the real or the risk-neutral measure. The QHR model can be used
either for modelling the actual price trajectories or for pricing
purpose. That distinction will be made also in the rest of the paper:
there will not be tilde-W Brownian motion. Conditional and forward
instantaneous variances will be the mathematical object. Neverthless,
it is interesting to investigates conditions where the same class of
QHR models can be used for both the real and the risk-neutral
measures. The following remark present one such condition.

\begin{remark}[A change of measure that preserves the model structure]
  Let
  $W_t = \tilde{W}_t + \int_0^t \frac{\mu_0 +
    \v\mu_1^T\v{y}_s}{\sigma_s} ds$ for some fixed $\mu_0 \in \Real$
  and $\v\mu_1 \in \Real^p$ such that
  $\v\mu_1^T\vv\Lambda^{-1}\v{b} \neq 1$. Assuming that $\tilde{W}_t$
  is a Brownian motion under a measure $\tilde{P}$ equivalent to $P$,
  the measure associated to the Brownian motion $W_t$. Then, $x_t$ and
  $\v{y}_t$ have dynamics
  \begin{subequations}\label{eq:mv:dynRW}
    \begin{align}
      dx_t &= (\mu_0 + \v\mu_1^T\v{y}_t - \frac12\sigma_t^2)dt + \sigma_td\tilde{W}_t,
      \\
      d\v{y}_t &= \mu_0\v{b} -(\vv\Lambda - \v{b}\v\mu_1^T)\v{y}_t dt + \v{b}\sigma_td\tilde{W}_t.
    \end{align}
  \end{subequations}
  Now, let $\tilde{\vv\Lambda} = \vv\Lambda - \v{b}\v\mu_1^T$ and
  $\tilde{\v{y}}_t = \v{y}_t - \mu_0\tilde{\vv\Lambda}^{-1}\v{b}$.
  That definition is well posed because
  $\v\mu_1^T\vv\Lambda^{-1}\v{b} \neq 1$ implies that
  $\tilde{\vv\Lambda}$ is non-singular. It follows that,
  \begin{align*}
    d\tilde{\v{y}}_t &= -\tilde{\vv\Lambda}\tilde{\v{y}}_t dt + \v{b}\sigma_tdW_t,
  \end{align*}
  and that $\sigma_t^2$ is a quadratic function of $\v{y}_t$. That is,
  the volatility process $\sigma_t$ (but not the underlying) follows a
  QHR model under the real world measure $\tilde{P}$. However, notice
  that, without specifying additional assumptions,
  $\tilde{\vv\Lambda}$ may have complex eigenvalues or negative
  eigenvalues. That is, the first condition of assumption \cref{h:1}
  may not be satisfied.

  The interpretation is the following, in the real world measure,
  $S_t$ can have a stochastic trend $\mu_t$ which is an affine
  function of the offset vector $\v{y}_t$:
  $dS_t = \mu_tS_tdt + \sigma_tS_tdW_t$.
  $\mu_t = \tilde\mu_0 + \tilde{\v\mu}_1^T\tilde{\v{y}}_t$ with
  $\tilde\mu_0\in \Real$ and $\tilde{\v\mu}_1 \in \Real^p$ fixed. \qed
\end{remark}

\subsection{Identification}
\label{sec:ident}

Let $\tilde{\v{y}}_t = \vv{M}^{-1}\v{y}_t$ for some non-singular
matrix $\vv{M} \in \Real^{p\times p}$. Then, \cref{eq:mv:dy} and
\cref{eq:mv:vol0} can be rewritten as
\begin{align*}
  d\tilde{\v{y}}_t
  &= - \tilde{\vv\Lambda}\tilde{\v{y}}_tdt
  + \tilde{\v{b}}\sigma_t dW_t,
  &
  \sigma_t^2
  &= \alpha
  + 2\tilde{\v{y}}_t^T\tilde{\v\beta}
  + \tilde{\v{y}}_t^T\tilde{\vv\Gamma}\tilde{\v{y}}_t,
\end{align*}
where $\tilde{\vv\Lambda} = \vv{M}^{-1}\vv\Lambda\vv{M}$,
$\tilde{\v{b}} = \vv{M}^{-1}\v{b}$,
$\tilde{\v\beta} = \vv{M}^T\v\beta$ and
$\tilde{\vv\Gamma} = \vv{M}^T\vv\Gamma\vv{M}$. That is, a change of
basis on the space of the latent process $\v{y}_t$ results in a model
having the same specification of \cref{eq:mv:dyn}. In other words,
model \cref{eq:mv:dyn} is not identified w.r.t. a similarity
transform applied to $\vv\Lambda$. To resolve that identification
issue, in the following it will be assumed that $\vv\Lambda$ has the
Jordan's structure
\begin{align}\label{eq:mv:LambdaJordan}
  \vv\Lambda &= \bigoplus_{i=1}^m \lambda_i \vv{D}_i,
  & 
  \vv{D}_i &= \pmx{
    1 & 0 & 0  & \cdots & 0 \\
    -1 & 1 & 0 & \ddots & \vdots \\
    0 & -1 & 1 & \ddots & 0 \\
    \vdots & \ddots & \ddots & \ddots & 0 \\
    0 & \cdots & 0 & - 1 & 1
  } \in \Real^{n_i\times n_i},
  &
  \lambda_1 > \cdots &> \lambda_m.
\end{align}

Notice that the assumption on distinct eigenvalues is necessary
because if $\vv\Lambda$ has two blocks having the identical
eigenvalue, then, corresponding to those two blocks, $\v{y}_t$ have
two sub-vectors that correspond to the same path integrals of
$\sigma_tdW_t$ and, consequently, have identical values. It follows
that, those two sub-vectors and their coefficients in the variance
specification can be aggregated and reducing the size of the model.

Now, even under the specification \cref{eq:mv:LambdaJordan}, the
model is still not identified w.r.t. the above similarity
transformations when $\vv{M} = \oplus_{i=1}^m \mu_i \eye_{n_i}$. In
order make the model identified w.r.t. those transformations, the
coefficient vector $\v{b}$ is assumed to be fixed and given by
\begin{align}\label{eq:mv:bIdentified}
  \v{b} &= \VEC(\{\v{b}_i\}_{i=1}^m),
  & \v{b}_i = \v{e}_1 \in \Real^{n_i}.
\end{align}

Clearly, a different set of constraints could have been chosen to
obtain identification. For instance, often, for continuous ARMA models
and for the COGARCH model, a companion matrix form is used
\cite{BrockwellChadraaLindner:2006}. The above Jordan structure has
been adopted because it simplifies the characterisation of its matrix
exponentials.

Resuming the following assumption will be made for the identification
of the QHR model.
\begin{assumption}[Identification - Jordan canonical form]\label{h:2}

  The matrix $\vv\Lambda$ and the vector $\v{b}$ have the structure
  given in \cref{eq:mv:LambdaJordan} and \cref{eq:mv:bIdentified}.
  
\end{assumption}

\subsection{Homogeneity properties}
\label{sec:homo}

We emphasise the following properties of the dynamics given in
\cref{eq:mv:dx} and \cref{eq:mv:dy}. This system of SDEs is
\begin{itemize}
\item Markovian,
\item autonomous (the coefficients do not depend on $t$) and
\item free of ``local volatility'' mechanisms.
\end{itemize}
For the last property we mean that the trend and diffusion
coefficients do no directly depend on the level of the log-price
$x_t$. 

One the main consequence of those properties is that, conditional on
$\mathcal{F}_t$, the distribution of the log-return $x_T - x_t$,
$T>t$, does not depend on $x_t$, but only on the offset process
$\v{y}_t$. Moreover, that distribution does not depend neither on the
absolute dates $t$ and $T$. More precisely, fix an horizon $\bar{T}>0$
and assume that the process $(x_t,\v{y}_t)$ admits a transition
density from $t$ to $T$, $0\leq t < T$. Then, calling
$p(T, x_T,\v{y}_T; t, x_t, \v{y}_t \;|\; \vv\Lambda, \alpha, \v\beta,
\vv\Gamma)$ that density can be written as
\begin{align*}
  p(T, x_T,\v{y}_T;  t, x_t, \v{y}_t
  \;|\;
  \vv\Lambda, \alpha, \v\beta, \vv\Gamma )
  &= 
  p_0(x_T-x_t,\v{y}_T; \v{y}_t, T-t
  \;|\;
  \vv\Lambda, \alpha, \v\beta, \vv\Gamma ),
\end{align*}
for some density
$p_0(\cdot,\cdot; \v{y}_t, T-t \;|\; \vv\Lambda, \alpha, \v\beta,
\vv\Gamma )$. Moreover, from \cref{eq:mv:dx},\cref{eq:mv:dy} and
\cref{eq:mv:vol0} the function $p_0$ satisfies the time-scaling
property
\begin{align*}
  p_0(x,\v{y}; \v{y}_0, cT \;|\;
  \vv\Lambda, \alpha, \v\beta, \vv\Gamma)
  &=
  p_0(x,\v{y}; \v{y}_0, T \;|\;
  c\vv\Lambda, c\alpha, c\v\beta, c\vv\Gamma),
  &
  c &>0.
\end{align*}

It follows that the price of a Call option,
\begin{align*}
  \text{Call}(S, K, \v{y}, t, T)
  &= \E[ (e^{x_T} - K)_+ \;|\; x_t=\log(S), \v{y}_t=\v{y}],
\end{align*}
is 0-homogeneous w.r.t. time and 1-homogeneous w.r.t. prices:
\begin{align*}
  \text{Call}(S, K, \v{y}, t, T)
  = S\,  \text{Call}(1, K/S, \v{y}; 0, T-t).
\end{align*}
By the Put-Call parity, Put prices have the same homogeneity
properties.

This property means that the option's relative price ($\text{Call}/S$)
depends only on the moneyness ($K/S$), on the current level of the
offsets ($\v{y}$) and on the time-to-maturity ($T-t$). In other words,
those relative price surface, ($K/S, T-t) \to \text{Call}/S$, depends
on the offsets and not of the absolute underlying price.

For options on spot variance, as \cref{eq:mv:dy} is Markovian, the
price depends only on the offset levels and time-to-maturity:
\begin{align*}
  \E[ \phi(\v\sigma_T^2) \;|\; \v{x}_t = x, \v{y}_t = \v{y}]
  &= 
  \E[ \phi(\v\sigma_{T-t}^2) \;|\; \v{x}_0 = 0, \v{y}_0 = \v{y}].
\end{align*}
Analogous considerations are valid for derivatives on the integrated
variance.

\subsection{Moving average filters}
\label{sec:filters}

Here, an heuristic characterisation of the latent factor $\v{y}_t$ as
moving averages of the price trajectories is given. The presentation
is only formal with heavy mathematical shortcuts.

Under \cref{h:1,h:2}, consider the moving average process defined in
\cref{eq:intro:tildey} and distinguish the two extreme cases of
distinct roots ($n_1 = n_2 = \cdots = n_m = 1$) and one single root
($m=1$, $n_1=n$) and then the general case.

\begin{itemize}
\item Distinct roots: $m\geq 1$, $p=m$ ($n_1=\cdots n_m=1$). 

  In this case, $\vv\Lambda$ is a diagonal matrix and $\v{b}$ is a
  vector of all ones:
  \begin{align*}
    \vv\Lambda &= \diag(\lambda_1,\ldots,\lambda_m),
    &
    \v{b} &= \ones.
  \end{align*}
  Then, $i$-th element of $\v{y}_t$ is given by
  \begin{align*}
    y_{it} &= \xi_t - \int_{-\infty}^t \lambda_i e^{-\lambda_i(t-s)} \xi_sds.
  \end{align*}
  That is, $y_{it}$ is the deviation of $\xi_t$ from a its exponential
  moving average with parameter $\lambda_i$. The moving average filter
  used to compute $y_{it}$ is the function
  $\phi_i: s\in \Real_+ \to \lambda_i e^{-\lambda_i s}$, which is strictly
  positive and integrates to one. Then, the application of that filter
  corresponds to on an average.

  Moreover, the elements of $\v{y}_t$ can be linearly combined to
  obtain another filter. For instance, if $\v{w} \in \Real_+^p$ and
  $\v{w}^T\v{b} = \v{w}^T\v\ones = 1$, then
  \begin{align*}
    \v{w}^T\v{y}_t &= \xi_t - \int_{\infty}^t \phi(t-s)\xi_s ds,
    &
    \phi(t) &= \sum_{i=1}^n w_i \lambda_i e^{-\lambda_i t}.
  \end{align*}
  Again, $\phi$ is positive and has unit integral.
  
\item Single root with multiplicity $n$: $m=1$, $n_1=n\geq 1$, $p=n$.

  In this case $\vv\Lambda = \lambda_1 \vv{D}$ and $\v{b} = \v{e}_1$.
  It turns out that
  \begin{align*}
    e^{-\vv\Lambda t} &=
    e^{-\lambda t}
    \pmx{
      1 & 0 & 0 & \cdots & 0 \\
      (\lambda t) & 1 & 0 & \ddots & \vdots  \\
      \frac{(\lambda t)^2}{2} & (\lambda t) & 1 & \ddots & 0 \\
      \vdots & \ddots & \ddots & \ddots & 0 \\
      \frac{ (\lambda t)^{n-1} }{(n-1)!}  & \cdots & \frac{(\lambda t)^2}{2} & (\lambda t) & 1
    }
  \end{align*}
  Then $e^{-\vv\Lambda t}\v{b}$ and, in turn,
  $\vv\Lambda e^{-\vv\Lambda t}\v{b}$ are, respectively, given by
  \begin{align*}
    e^{-\vv\Lambda t}\v{b} &= e^{-\lambda t}\pmx{
      1 \\ (\lambda t) \\  \frac{(\lambda t)^2}{2} \\
      \vdots \\   \frac{ (\lambda t)^{n-1} }{(n-1)!}
    }
    &\text{and}&&
    \vv\Lambda e^{-\vv\Lambda t}\v{b} &=
    \lambda e^{-\lambda t}
    \pmx{
      1 \\
      \lambda t -1 \\
      \frac{(\lambda t)^2}{2} - \lambda t \\
      \vdots \\
      \frac{ (\lambda t)^{n-1} }{(n-1)!} - 
      \frac{ (\lambda t)^{n-2} }{(n-2)!}
    }.
  \end{align*}
  The first element of $\v{y}_t$ is given by the offset
  \begin{subequations}
  \begin{align}\label{eq:mv:y1}
    y_{1t} &= \xi_t - \int_{-\infty}^t \psi_1(t-s) \xi_s ds,
    \intertext{and the remaining ones by the offsets}
    \label{eq:mv:yi}
    y_{it}
    &= \int_{-\infty}^t \psi_i(t-s)\xi_s ds
    - \int_{-\infty}^t \psi_{i-1}(t-s) \xi_s ds,
    &
    i &>1.
  \end{align}    
  \end{subequations}
  where
  $ \psi_i(t) =
  \lambda e^{-\lambda t}\frac{(\lambda t)^{i-1}}{(i-1)!}$, $1 \leq i \leq n$.
  Since $\lambda > 0$, $\int_0^\infty\psi_i(t)dt = 1$ for all $i$.
  It turns out that the filter $\phi$ defined in
  \cref{eq:intro:tildey} can be written as
  \begin{align*}
    \phi(t)
    =
    \sum_{i=1}^{n-1} (w_i - w_{i+1}) \psi_i(t)
    + w_n \psi_n(t).
  \end{align*}
  Then, if $w_1 = 1$ and $w_i \geq w_{i+1} \geq 0$, $i<n$, the function
  $\phi$ is positive and integrates to one.
  \Cref{fig:mv:filt0} shows the functions $\phi_i$,
  $i=1,\ldots,5$ when $\lambda=4$ and one example of $\psi(t)$ when
  $\v{w} = (1; 0.8; 0.6; 0.4)$. Note that, while the filter $\psi_1$
  corresponds to an exponential moving average, $\psi_2, \psi_3$ and
  $\psi_4$ correspond to moving averages that concentrate more weight
  near a lag of 0.25, 0.5 and 0.75 years. The processes $y_{2t}$,
  $y_{3t}$ and $y_{4t}$ are offsets between those moving average.
  Moreover, $\phi$ is an average of the four filters
  $\psi_1,\ldots, \psi_4$:
  $\phi = 0.2 \psi_1 + 0.2 \psi_2 + 0.2 \psi_3 + 0.4 \psi_4$.
  \begin{figure}[htb]
    \centering

    \includegraphics[scale=0.58]{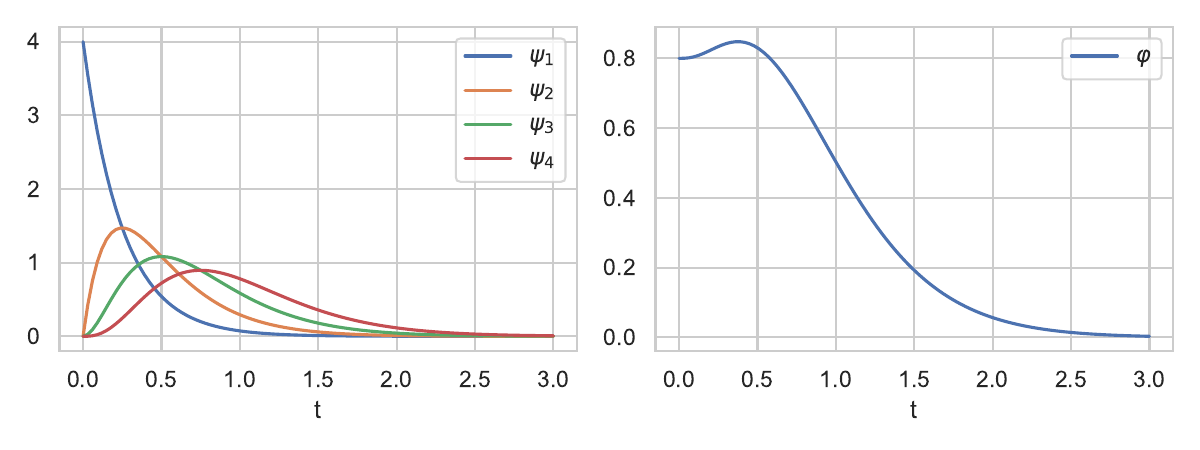}
    
    \caption{Left plot: the filters $\psi_1(t),\ldots,\psi_4(t)$ when
      $\lambda=4$. Right plot: the moving average filter $\phi(t)$
      when $\lambda=4$ and $\v{w}^T = \pmx{1 & 0.8 & 0.6 & 0.4}$. }
    \label{fig:mv:filt0}
  \end{figure}

\item $m>1$, $n_i \geq 1$, $i=1,\ldots,m$.

  In this case $\vv\Lambda$ is a block diagonal matrix with the $i$-th
  block being $\lambda_i \vv{D}$. Moreover, $\v{b}$ is a vector with
  blocks given by $\v{e}_1$. Then $\v{y}_t$ can be partitioned into
  $m$ blocks of dimension $n_1,\ldots,n_m$, and for each blocks are
  valid the considerations done for the previous case.

  To obtain a positive moving average filter as linear combination of
  the elements of $\v{y}_t$ it is sufficient to choose
  $\v{w} \in \Real^p$ such that $\v{w}^T\v{b} = 1$ and, partitioning
  $\v{w}$ into blocks of dimension $n_1,\ldots,n_m$, the elements of
  each block in decreasing order with the first one being positive.
  Note that $\v{w}^T\v{b}$ is the sum of the first elements of those
  blocks.
  
\end{itemize}

\medskip

With the interpretation of the elements $\v{y}_t$, or their linear
combinations, as deviations of $\xi_t$ from different moving averages,
it turns out that the instantaneous variance is a quadratic function
of those offset processes. Let assume that we want $\sigma_t^2$ to be
a quadratic function of the set of offsets
$\v{w}_1^T \v{y}_t, \ldots, \v{w}_r^T\v{y}_t$, for some vectors
$\v{w}_1, \ldots, \v{w}_r \in \Real^p$, $1 \leq r$. Then,
\begin{align}\label{eq:mv:vol2}
  \sigma_t^2 &= \alpha
  + 2\v\beta_0^T \vv{W}^T\v{y}_t
  + \v{y}_t^T \vv{W}\vv\Gamma_0\vv{W}^T\v{y}_t,
\end{align}
for some $\alpha > 0$, $\v\beta_0 \in \Real^r$ and
$\vv\Gamma_0 \in \mathcal{S}_+^r$, and where
$\vv{W} \in \Real^{p \times r}$ is the matrix with columns
$\v{w}_1, \ldots, \v{w}_r$.
This is exactly the model presented in \cref{eq:mv:vol0}, where
$\v\beta = \vv{W}\v\beta_0$ and
$\v\Gamma = \vv{W}\vv\Gamma_0\vv{W}^T$.

\subsection{Conditional moments}
\label{sec:conditionalMoments}

In the following, additional assumptions will be imposed to guarantee
that the first two conditional moments of $\sigma_t^2$ do not diverge
as $t\to \infty$. This property implies that the model allows for
weak-stationarity of the variance process.

To this aim, rewrite the variance function as
\begin{align}\label{eq:mv:vol1}
  \sigma_t^2 = \alpha + 2\v\beta^T\v{y}_t + \v\gamma^T\v{q}_t,
\end{align}
where $\v\gamma = \VEC(\vv\Gamma)$ and
$\v{q}_t = (\v{y}_t \otimes \v{y}_t)$. Then,
\begin{align*}
  \E[\v\sigma_t^2 \;|\; \mathcal{F}_0]
  &= \alpha + 2\v\beta^T \v{m}_0^{(1)}(t) + \v\gamma^T \v{m}_0^{(2)}(t),
\end{align*}
where $\v{m}_0^{(1)}(t) = \E[ \v{y}_t \;|\; \mathcal{F}_0]$ and
$\v{m}_0^{(2)}(t) = \E[ \v{q}_t \;|\; \mathcal{F}_0]$. The following
result provides a characterisation of the conditional first four
moments of $\v{y}_t$, or, equivalently, of the first two moments of
$(\v{y}_t; \v{q}_t)$.

The following Lemma characterises the conditional moments and marginal
moments 
\begin{align*}
  \v{m}_0^{(k)}(t) &= \E[ \v{y}_{t+s}^{\otimes k} \;|\; \mathcal{F}_0],
  & 
  \v{m}^{(k)}(t) &= \E[ \v{y}_{t+s}^{\otimes k}],
  &
  k &= 1,\ldots,
\end{align*}
as functions of $t \in [0,T-s]$ for a fixed $s\geq 0$. Next, a
corollary provides an analogous result for the marginal moments
$\v{m}^{(k)}(t) = \E[\v{y}^{\otimes k}]$. In the following, the
derivative of $f(t)$ w.r.t. its arguments is represented by
$\dot{f}(t) = df(t) / dt$.

\begin{lemma}\label{thm:mv:momentODE}
  If \cref{h:1} holds, then
  $\v{m}_0 = (\v{m}_0^{(1)}; \v{m}_0^{(2)}; \v{m}_0^{(3)};
  \v{m}_0^{(4)})$ is solution to the lower block-triangular ODE
  \begin{align}\label{eq:mv:muODE}
    \underbrace{
      \pmx{
        \dot{\v{m}}_0^{(1)} \\  \dot{\v{m}}_0^{(2)} \\
        \dot{\v{m}}_0^{(3)} \\  \dot{\v{m}}_0^{(4)}
      }
    }_{\dot{\v{m}}_0}
    &=
    \underbrace{
      \pmx{0 \\ \alpha \bar{\v{b}} \\ 0 \\ 0}
    }_{\v{a}}
    -
    \underbrace{
      \pmx{
        \vv{A}_{11} & 0 & 0 & 0 \\
        \vv{A}_{21} & \vv{A}_{22} & 0 & 0 \\
        \vv{A}_{31} & \vv{A}_{32} & \vv{A}_{33} & 0 \\
        0 & \vv{A}_{42} & \vv{A}_{43} & \vv{A}_{44}
      }
    }_{\vv{A}}
    \underbrace{
      \pmx{
        \v{m}_0^{(1)} \\  \v{m}_0^{(2)} \\  \v{m}_0^{(3)} \\  \v{m}_0^{(4)}
      }
    }_{\v{m}_0},
    &&
    \text{on } [0,T],
    \intertext{with initial condition}
    \label{eq:mv:condODE0}
    \v{m}_0^{(k)}(0) &= \v{y}_0^{\otimes k},
    &&
    k = 1,2,3,4, 
  \end{align}
  where $\bar{\v{b}} = \v{b}\otimes \v{b}$ and the
  blocks of $\vv{A}$ are given by
  \begin{align}\label{eq:mv:Ak}
    \vv{A}_{kk} &= \vv\Lambda^{(k)} - \vv{B}^{(k)}\otimes \v\gamma^T, 
    &
    \vv{A}_{k,k-1} &= -2\vv{B}^{(k)}\otimes \v\beta^T, 
    &
    \vv{A}_{k,k-2} &= -\alpha\vv{B}^{(k)},
  \end{align}
  with $\vv\Lambda^{(k)}$, $\vv{B}^{(k)}$ and $\vv{C}^{(k)}$
  recursively given by
  \begin{subequations}\label{eq:mv:LBCk}
    \begin{align}
      \label{eq:mv:Lk}
      \vv\Lambda^{(1)} &= \vv\Lambda,
      &&&
      \vv\Lambda^{(k+1)} &= \eye_p \otimes \vv\Lambda^{(k)} + \vv\Lambda\otimes \eye_{p^k},
      & k\geq 1,
      \\
      \label{eq:mv:Bk}
      \vv{B}^{(1)} &= 0,
      &&&
      \vv{B}^{(k+1)} &= \eye_p \otimes \vv{B}^{(k)} + \v{b} \otimes \vv{C}^{(k)},
      & k\geq 1,
      \\
      \label{eq:mv:Ck}
      \vv{C}^{(1)} &= \v{b},
      &&&
      \vv{C}^{(k+1)} &= \eye_p \otimes \vv{C}^{(k)} + \v{b} \otimes \eye_{p^k}.
      & k\geq 1,
    \end{align}
  \end{subequations}
  Moreover,
  $\v{m} = (\v{m}^{(1)}; \v{m}^{(2)}; \v{m}^{(3)}; \v{m}^{(4)})$
  solves the ODE \cref{eq:mv:muODE} with initial condition
  $\v{m}^{(k)}(0) = \E[ \v{y}_0^{\otimes k}]$, $k=1,2,3,4$.

  When $\vv{A}$ is non-singular,
  \begin{align*}
    \v{m}_0(t) &= (\eye - e^{-\vv{A}t})\v{m}_{\infty} + e^{-\vv{A}t}\v{m}_0(0),
    \intertext{and}
    \v{m}(t) &= (\eye - e^{-\vv{A}t})\v{m}_{\infty} + e^{-\vv{A}t}\v{m}(0),
  \end{align*}
  where $\v{m}_{\infty} = \vv{A}^{-1}\v{a}$.
  
  \begin{proof}
    See \cref{sec:proofs}.
  \end{proof}
\end{lemma}

\medskip

Regarding the structure of the matrix $\vv{A}$ defined in
\cref{eq:mv:muODE}, note that $\bar{\v{b}} = \vv{B}^{(2)}$,
$ \bar{\vv\Lambda} = \vv\Lambda^{(2)} = \vv\Lambda \otimes \eye_p +
\eye_p \otimes \vv\Lambda$ and the top-left sub-matrix of $\vv{A}$ is
given by
\begin{align}\label{eq:mv:tildeA}
  \tilde{\vv{A}}
  &=
  \pmx{
    \vv{A}_{11} & 0 \\ \vv{A}_{21} & \vv{A}_{22}    
  }
  =
  \pmx{
    \vv\Lambda & 0 \\ 
    - 2\bar{\v{b}}\v\beta^T & \bar{\vv\Lambda} - \bar{\v{b}}\v\gamma^T
  }.
\end{align}
Moreover, from the structure of $\vv{A}$ and $\v{a}$ it follows that
the first block of $\v{m}_\infty$ is zero ($\v{m}_\infty^{(1)} = 0$)
and the second block, hereafter denoted by $\v{q}_\infty$, is given by
\begin{align}\label{eq:mv:qinf}
  \v{q}_\infty = \v{m}_{\infty}^{(2)}
  &=
  \alpha\vv{A}_{22}^{-1}\bar{\v{b}}
  = (1-\kappa)^{-1}\alpha \bar{\vv\Lambda}^{-1}\bar{\v{b}},
  &
  \kappa &= \v\gamma^T\bar{\vv\Lambda}^{-1}\bar{\v{b}}.
\end{align}

Since the coefficient matrix $\vv{A}$ is block triangular, the
stability of the ODE \cref{eq:mv:muODE} is determined by the
eigenvalues of the diagonal blocks. More precisely, that ODE is stable
when all the eigenvalues of $\vv{A}_{22}$, $\vv{A}_{33}$ and
$\vv{A}_{44}$ have positive real part. In the following Lemma a
sufficient condition for that stability is provided.

\begin{lemma}[Stability of  models with non-negative hessian $\vv\Gamma$]
  \label{thm:mv:stability}

  Under \cref{h:1,h:2}, if 
  \begin{align}\label{eq:mv:tkappa}
    \vv\Gamma &\geq 0
    &&\text{and}&
    \tilde\kappa &=
    \lambda_{\min}
    \v{b}^T\vv\Lambda^{-T}\vv\Gamma\vv\Lambda^{-1}\v{b}
    < \frac23,
  \end{align}
  then all the eigenvalues of $\vv{A}$ have positive real part.

  \begin{proof}
    See \cref{sec:proofs}.
  \end{proof}
\end{lemma}

\subsection{Autocovariance function and Weak stationarity of the
  variance process}
\label{sec:weakStationarity}

When all the eigenvalues of $\vv{A}$ have positive real part the ODE
is stable, and one can consider to expand the horizon of the
solutions: $T \to \infty$. In that case,
$\lim_{t\to \infty} \v{m}_0(t) = \v{m}_{\infty}$. Moreover, if the
distribution of $\v{y}_0$ has the first two moments given by
$\v{m}^{(1)}_{\infty}$ and $\v{m}^{(2)}_\infty$, then the process
$\v{y}_t$ will be weak-stationary. Being a linear function of
$\v{y}_t$ and $\v{y}_t^{\otimes 2}$, the variance process
$\sigma_t^2$ is weak-stationary when the moments up to the fourth
order of $\v{y}_0$ are given by $\v{m}_\infty$.

The following assumption is related to the stability of the SDE
\cref{eq:mv:muODE} and to the weak-stationarity of both the offset
process $\v{y}_t$ and the variance process $\sigma_t^2$.

\begin{assumption}[Weak stationarity of the variance process]
  \label{h:3}

  The eigenvalues of $\vv{A}_{kk}$, $1\leq k\leq 4$ have strictly
  positive real parts and the first fourth-order moments of $\v{y}_0$
  are given by $\v{m}_\infty = \vv{A}^{-1}\v{a}$.
\end{assumption}

Consider the $p + p^2$ dimensional process
$\v\eta_t = (\v{y}_t; \v{q}_t)$, $\v{q}_t = \v{y}_t\otimes \v{y}_t$.
Clearly, the first four moments of $\v{y}_t$ correspond to the first
two moments of $ \v\eta_t$ and, thus, $\v{m}(t)$ contains those
moments. It follows that, under \cref{h:1,h:3}, since
$\v{m}(0) = \v{m}_\infty$, the first two moments of $\v\eta_t$ have
mean and variance that do not depend on $t$:
\begin{align}\label{eq:mv:momEta}
  \E[ \v\eta ] &= \v\eta_\infty = \pmx{ 0 \\ \v{q}_\infty},
  &\text{and}&&
  \E\big[ \v\eta_t\v\eta_t^T \big]
  &=
  \pmx{ \vv{M}_2 & \vv{M}_3 \\ \vv{M}_3^T & \vv{M}_4 },
\end{align}
where $\vv{M}_2 \in \Real^{p\times p}$,
$\vv{M}_3 \in \Real^{p \times p^2}$ and
$\vv{M}_4 \in \Real^{p^2 \times p^2}$ are such that
$\VEC(\vv{M}_k) = \v{m}^{(k)}_\infty$, $k=2,3,4$. It follows that the
covariance matrix of $\v\eta_t$ is
\begin{align}\label{eq:mv:OmegaDef}
  \Cov(\v\eta_t)
  &=
  \vv\Omega = 
  \pmx{
    \vv{M}_2 & \vv{M}_3 \\
    \vv{M}_3^T & \vv{M}_4 - \v{q}_\infty \v{q}_\infty^T
  }.
\end{align}
We will not continue to seek an explicit an explicit expression for
$\vv{M}_3$ and $\vv{M}_4$, as these are more complex to implement than
simply solving the block-triangular system
$\vv{A}\v{m}_\infty = \v{a}$ numerically.

\begin{proposition}[Conditional expectation]
  Under \cref{h:1}, if $\vv{A}$ is non-singular, then
  \begin{align}\label{eq:mv:condMomentEta}
    \E[\v\eta_{t+s}\;|\;\mathcal{F}_t]
    &= \v\eta_\infty + e^{-\tilde{\vv{A}}s}(\v\eta_s - \v\eta_\infty),
  \end{align}
  where $\v\eta_\infty$ and $\tilde{\vv{A}}$ are defined in
  \cref{eq:mv:momEta} and \cref{eq:mv:tildeA}, respectively.
  Moreover, the conditional instantaneous variance is given by
  \begin{align}\label{eq:mv:fwdVar}
    v_t(s)
    &= \E[ \sigma_{t+s}^2 \;|\; \mathcal{F}_t]
    = \sigma_{\infty}^2
    + \v\psi(s)^T (\v\eta_t - \v\eta_\infty),
  \end{align}
  where
  \begin{align}\label{eq:mv:varInf}
    \sigma_\infty^2 &= \frac\alpha{1-\kappa},
    &
    \v\psi(s) &= (e^{-\tilde{\vv{A}}s})^T\v{g},
    &
    \v{g}^T = (2\v\beta^T\; \v\gamma^T),
  \end{align}
  and $\kappa$ is defined in \cref{eq:mv:qinf}.

  \begin{proof}
    The first result is a direct consequence of 
    \cref{thm:mv:momentODE}. The second one, follows by rewriting the
    variance as an affine combination of $\v\eta_t$:
    $\sigma_t^2 = \alpha + \v{g}^T\v\eta_t$.
  \end{proof}
\end{proposition}

\begin{lemma}[Weak stationarity of $\v\eta_t$ and $\sigma_t^2$]
  If \cref{h:1,h:3} hold, then the process $\v\eta_t$ is
  weak-stationary with expectation $\v\eta_\infty$ and covariance
  matrix $\vv\Omega$ given, respectively, in \cref{eq:mv:momEta} and
  and auto-covariance function
  \begin{align}\label{eq:mv:autocovEta}
    s \to \Cov(\v\eta_{t+s},\v\eta_t^T) = e^{-\tilde{\vv{A}}s}\vv\Omega,
  \end{align}
  where $\tilde{\vv{A}}$ is defined in \cref{eq:mv:tildeA}.
  
  Moreover, the process $\sigma_t^2$ is weak-stationary with
  \begin{align*}
    \E[\sigma_t^2]
    &= \sigma_\infty^2,
    &\text{and}&&
    \E[ \sigma_t^4 ]
    &= \alpha
    + \v{g}^T \vv\Omega \v{g},
  \end{align*}
  and autocovariance function given by
  \begin{align}\label{eq:mv:autocovVar}
    s \to \Cov(\sigma^2_{t+s}, \sigma^2_t)
    =
    \v\psi(s)^T \vv\Omega \v{g},
  \end{align}
  where $\sigma^2_\infty$, $\v\psi(s)$ and $\v{g}$ are defined in
  \cref{eq:mv:varInf}.

  \begin{proof}
    Note that the first two order moments of $\v\eta_t$ corresponds to
    the first four order moments of $\v{y}_t$. Then, since from
    \cref{h:3} all the eigenvalues of $\vv{A}$ have positive real
    part, the weak stationarity of $\v\eta_t$ is a direct consequence
    of \cref{thm:mv:momentODE}.

    Next, to obtain \cref{eq:mv:autocovEta}, apply the decomposition of
    variance rule and \cref{eq:mv:fwdVar}:
    \begin{align*}
      \Cov( \v\eta_{t+s}, \v\eta_t^T)
      &=
      \underbrace{
        \Cov\!\big( \E[ \v\eta_{t+s}\;|\; \mathcal{F}_t], \v\eta_t \big)
      }_{
        = \Cov( e^{-\tilde{\vv{A}}s}\v\eta_t, \v\eta_t)
      }
      +
      \E\!\big[
        \underbrace{
          \Cov( \v\eta_{t+s}, \v\eta_t^T \;|\; \mathcal{F}_t) 
        }_{=0}
        \big]
      = e^{-\tilde{\vv{A}}s}\vv\Omega.
    \end{align*}
    The remaining results follow by recalling that
    $\sigma_t^2 = \alpha + \v{g}^T\v\eta_t$.
    
  \end{proof}
\end{lemma}

\bigskip

Being $\sigma_\infty^2$
the log-return's variance, the corresponding Kurtosis is defined as
\begin{align}\label{eq:mv:kurtosis}
  \text{Kurt}_\infty = \E[\sigma_t^4] / \sigma_{\infty}^4.
\end{align}

\begin{remark}
  One implication of these considerations is that, although the
  dynamics specification of $\v{y}_t$ involve only the $m$ roots
  $\lambda_1, \ldots, \lambda_m$, the feedback due to dependence of
  the volatility coefficient on $\v{y}_t$ induces additional
  artificial roots on the auto-correlation structure of the processes
  $\v{y}_t$ and, in turn, $\v\sigma_t^2$.
\end{remark}

\subsection{Forward variance term structure and its principal
  component decomposition}
\label{sec:fwdVariance}

From a financial point of view, when the SDEs
\cref{eq:mv:dx}-\cref{eq:mv:dy} correspond to the discounted
log-price's risk-neutral dynamics, the conditional variance $v_t(s)$
represents the ``forward instantaneous variance'', or simply ``forward
variance'', for the maturity $t+s$. Note that, seen as function of
$\v{y}_t$, $v_t(s)$ is quadratic, while seen as function of
$\v\eta_t$, $v_t$ is affine \cite{GatheralKellerRessel:2019}. Then,
the forward variance curve can be decomposed into a fixed term, a
linear combination of the factors $\v{y}_t$ and one of their
``interactions'' $\v{q}_t = \v{y}_t \otimes \v{y}_t$:
\begin{multline}\label{eq:mv:fwdVarDecomposition}
  v_t(s)
  = v^0(s)
  + \sum_{i=1}^p \psi_i^{(\v{y})}(s) y_{it}
  + \sum_{i,j=1}^p \psi_{i,j}^{(\vv{Q})}(s) y_{it}y_{jt}
  \\
  = v^0(s)
  + \v{y}_t^T\v\psi^{(\v{y})}(s)
  + \v{y}_t^T  \vv\Psi^{(\vv{Q})}(s) \v{y}_t.
\end{multline}
Here, $\v\psi^{(\v{y})}$ corresponds to the first $p$ elements of
$\v\psi$, while $\vv\Psi^{(Q)}$ to the last $p^2$. Clearly, as the
coefficients of the quadratic function $\v{y}_t \to v_t(s)$ are
functions of the time-to-maturity $s$, also its minimum depend on $s$.
More precisely,
\begin{align}\label{eq:mv:fwdVarMin}
  v^{\min}(s) = \min_{\v{y}_t} v_t(s)
  &=
  v^0(s)
  + \frac14 \v\psi^{(\v{y})}(s)^T
  \big( \vv\Psi^{(\vv{Q})}(s) \big)^{-1}
  \v\psi^{(\v{y})}(s),
  &
  s, t &\geq 0.
\end{align}
Moreover, since $v_t(s) \to v_\infty$ as $s\to \infty$, that convex
function of $\v{y}_t$ asymptotically flattens to the stationary level.
Some instances of the curves $v^0$ and $v^{\min}$ are shown in \cref{fig:r1:pca}.

Since the decomposition in \cref{eq:mv:fwdVarDecomposition} depends
on the parameterisation, it is be more appropriate to consider a
principal component decomposition that depend only on the first two
moments of the forward variance curve $v_t(s)$. The following
proposition describes an algorithm for obtaining the factors of such
decomposition.

\begin{proposition}[Principal Component decomposition of the Forward Variance curve]
  If \cref{h:1,h:3} hold, then the principal component decomposition
  of the forward variance curve $v_t(\cdot)$ is given by
  \begin{subequations}
    \label{eq:mv:fwdVarPCA}
    \begin{align}
      \Cov( v_t(s_1), v_t(s_2) )
      &= \v{u}(s_1)^T\vv{D} \v{u}(s_2),
      &
      t,s_1,s_2 &\geq 0,
      \\
      \int_0^\infty \v{u}(t)\v{u}(t)^Tdt &= \eye.
    \end{align}
  \end{subequations}
  where the diagonal matrix with the eigenvalues $\diag(\vv{D})$ and
  the principal factors vector $\v{u}(\cdot)$ are defined as follows.

  Let $\vv{R} \in \Real^{(p+p^2) \times r}$ be a Cholesky factor of
  \begin{align}\label{eq:mv:Fdef}
    \vv{F}
    &= \int_0^\infty \v\psi(t) \v\psi(t)^T dt = \vv{R}\vv{R}^T,
  \end{align}
  where $1 \leq r \leq (p+p^2)$ is the rank of $\vv{F}$ and $\vv{R}$
  has full rank. Next, consider the eigenvalue decomposition of
  $\vv{R}^T\vv\Omega\vv{R}$, with $\vv\Omega$ being the covariance
  matrix defined in \cref{eq:mv:OmegaDef}:
  \begin{align*}
    \vv{R}^T\vv\Omega\vv{R} = \vv{V}\vv{D}\vv{V}^T,
  \end{align*}
  where $\vv{V}, \vv{D} \in \Real^{r \times r}$,
  $\vv{V}$ orthogonal and $\vv{D} \geq 0$ diagonal. Finally, define
  $\v{u}:\Real_+ \to \Real^r$ as
  \begin{align*}
    \v{u}(t) = \vv{V}^T\vv{R}^{+}\v\psi(t),
  \end{align*}
  where $\vv{R}^{+} \in \Real^{r \times (p + p^2)}$ is a
  pseudo-inverse of $\vv{R}$ such that $\vv{R}^+\vv{R} = \eye_r$.

  \begin{proof}
    Firstly, by \cref{h:1,h:3} all the eigenvalues of $\tilde{\vv{A}}$
    have positive real part, and, since
    $\v\psi(t) = (e^{-\tilde{\vv{A}}s})^T\v{g}$, the integral in
    \cref{eq:mv:Fdef} is finite. Moreover, the covariance matrix
    $\vv\Omega$ is well defined. Now, since the integrand in
    \cref{eq:mv:Fdef} is positive semidefinite, it is necessary that
    $\psi(t)$ belong to the column range of $\vv{R}$ for all
    $t\geq 0$. It follows that $\v\psi(t) = \vv{R}\vv{V}\v{u}(t)$.
    Finally, by \cref{eq:mv:fwdVar} and \cref{eq:mv:OmegaDef}, for
    $t,s_1,s_2\geq 0$,
    \begin{align*}
      \Cov(v_t(s_1),v_t(s_2))
      &= \v\psi(s_1)^T\vv\Omega \v\psi(s_2)
      = \v{u}(s_1)^T\vv{V}^T\vv{R} \vv\Omega \v{R}\vv{V}\v{u}(s_2)
      = \v{u}(s_1)^T\vv{D}\v{u}(s_2).
    \end{align*}
    
  \end{proof}
\end{proposition}

\begin{remark}
  Note that, $\v\eta_t = (\v{y}_t; \VEC(\vv{Q}_t))$ where
  $\vv{Q}_t = \v{y}_t\v{y}^T$ is symmetric. Then, $p(p-1)$ elements of
  $\v\eta_t$ have an identical copy. It follows that the covariance
  matrix $\vv\Omega$ has $p(p-1)/2$ null eigenvalues, and the number
  of factors in the PCA \cref{eq:mv:fwdVarPCA} can be reduced to
  $p(p+3)/2$ without any loss of information. \qed
\end{remark}

\begin{remark}
  The integral in \cref{eq:mv:Fdef} can be explicitly computed in
  terms of the matrix $\tilde{\vv{A}}$ as follows. Since
  $\v\psi(t) = (e^{-\tilde{\vv{A}}t})^T\v{g}$, the integrand can be
  rewritten as
  \begin{align*}
    \VEC(\v\psi(t) \v\psi(t)^T)
    &=
    \v\psi(t) \otimes \v\psi(t)
    = 
    (e^{-\tilde{\vv{A}}t} \otimes e^{-\tilde{\vv{A}}t})^T
    (\v{g} \otimes \v{g})
    = 
    (e^{-(\tilde{\vv{A}} \otimes \eye + \eye \otimes \tilde{\vv{A}})t})^T
    (\v{g} \otimes \v{g}),
  \end{align*}
  and, thus,
  \begin{align*}
    \VEC(F)
    % &= \int_0^\infty \VEC(\v\psi(t)\v\psi(t)^T)dt
    % \\
    &= 
    - \Big[
      (\tilde{\vv{A}} \otimes \eye + \eye \otimes \tilde{\vv{A}})^{-1}
      (e^{-(\tilde{\vv{A}} \otimes \eye + \eye \otimes \tilde{\vv{A}})t})^T
      (\v{g}\otimes \v{g})
      \Big]_0^\infty
    \\
    &= 
    (\tilde{\vv{A}} \otimes \eye + \eye \otimes \tilde{\vv{A}})^{-1}
    (\v{g}\otimes \v{g}).
  \end{align*}
  That is, $\vv{F}$ solves the matrix equation
  $\tilde{\vv{A}}^T\vv{F} + \vv{F}\tilde{\vv{A}} = \v{g}\v{g}^T$.
  \qed
\end{remark}

\subsection{The autocorrelation of squared increments}
\label{sec:squaredIncrements}

Following \cite{BrockwellChadraaLindner:2006}, consider the increment
process
\begin{align*}
  \xi_t^{(r)} &= \xi_{t+r} - \xi_t,
  &
  \xi_t &= \int_0^t \sigma_s dW_s.
\end{align*}

The following Lemma characterises the autocorrelation structure of the
increment and squared increment processes $\xi_t^{(r)}$ and
$(\xi_t^{(r)})^2$.

\begin{lemma}
  If \cref{h:1,h:3} hold, then for any $t\geq 0$ and
  $h \geq r\geq 0$,
  \begin{align*}
    \E[\xi_t^{(r)}] &= 0,
    &
    \Cov( \xi_t^{(r)}, \xi_{t+h}^{(r)}) &= 0,
    \\
    \E\!\big[ (\xi_t^{(r)})^2 \big] &= r \sigma_\infty^2,
    &
    \Cov\!\big( (\xi_t^{(r)})^2, (\xi_{t+h}^{(r)})^2 \big)
    &=
    \v{g}^Te^{-\tilde{\vv{A}}h}\v{h}_r,
  \end{align*}
  where
  \begin{align*}
    \v{h}_r &=
    \tilde{\vv{A}}^{-1}(e^{\tilde{\vv{A}}r}- \eye)
    \Cov(\v\eta_r, \xi_r^2).
  \end{align*}

  \begin{proof}
    Without loss of generality assume $t=0$. The null expectation of
    $\xi_0^{(r)}$ is a direct consequence of its definition. Consider
    its variance,
    \begin{align*}
      \E[ (\xi_0^{(r)})^2]
      = \E[ \xi_r^2 ] = \E\Big[\Big( \int_0^r \sigma_s dW_s\Big)^2\Big]
      = r \sigma_\infty^2,
    \end{align*}
    since the process $\sigma_s^2$ is stationary with expectation
    $\sigma_\infty^2$.
    Next,
    \begin{align*}
      \Cov( \xi_0^{(r)}, \xi_h^{(r)})
      &= \E[ \xi_0^{(r)}\xi_h^{(r)}]
      = \E\big[ \xi_0^{(r)} \E[ \xi_h^{(r)} \;|\; \mathcal{F}_h] \big]
      = 0.
    \end{align*}
    Finally, to obtain the expression for the autocovariance function
    of squared increments, note that by \cref{eq:mv:fwdVar}
    \begin{align*}
      \E[ (\xi_h^{(r)})^2 \;|\; \mathcal{F}_r]
      &= 
      \E\Big[ \int_h^{h+r} \sigma_s^2 ds \;\Big|\; \mathcal{F}_r\Big]
      = \int_{h-r}^h v_r(s) ds
      \\
      &=
      r\sigma_\infty^2
      + \int_{h-r}^h \v\psi(s)^T(\v\eta_r - \v\eta_\infty) ds
      \\
      &= 
      r\sigma_\infty^2
      +
      \v{g}^Te^{-\tilde{\vv{A}}h}
      \tilde{\vv{A}}^{-1}\big( e^{\tilde{\vv{A}}r}-\eye \big)
      (\v\eta_r - \v\eta_\infty).
    \end{align*}
    Now, since $\xi_0^r=\xi_r$,
    $\E[ (\xi_0^{(r)})^2 (\xi_h^{(r)})^2]
    = \E[ \xi_r^2 (\xi_h^{(r)})^2]$
    and thus
      
    \begin{align*}
      \E[ (\xi_0^{(r)})^2 (\xi_h^{(r)})^2]
      &=
      \E\!\Big[
        \xi_r^2
        \E[ (\xi_h^{(r)})^2 \;|\; \mathcal{F}_r]
        \Big]
      \\
      &= 
      r\sigma_\infty^2\E[ \xi_r^2]
      +
      \v{g}^Te^{-\tilde{\vv{A}}h}
      \tilde{\vv{A}}^{-1}\big( e^{\tilde{\vv{A}}r}-\eye \big)
      \E[\xi_r^2 (\v\eta_r - \v\eta_\infty) ]
      \\
      &=
      (r\sigma_\infty^2)^2
      +
      \v{g}^Te^{-\tilde{\vv{A}}h}
      \tilde{\vv{A}}^{-1}\big( e^{\tilde{\vv{A}}r}-\eye \big)
      \Cov(\v\eta_r, \xi_r^2).
    \end{align*}    
  \end{proof}
\end{lemma}

\section{Specific models and numerical tests}
\label{sec:tests}
\subsection{Scalar QHR model}

When $p=m=n=1$, the SDE \cref{eq:mv:dy} can be rewritten as
\begin{align}\label{eq:scalar:dy}
  dy_t &= - \lambda y_tdt + \sigma_t dW_t,
  &
  \sigma_t^2 &= \alpha + 2\beta y_t + \gamma y_t^2,
\end{align}
where $\lambda,\alpha >0$ and $\beta,\gamma \in \Real$ are such that
$\gamma\alpha > \beta^2$. The variance function is quadratic function
whose minimum is located at $y_{\min} = -\gamma/\beta$ and has value
$\sigma_{\min}^2 = \alpha^2 - \gamma/\beta^2$.

The variance is function of the offset from an exponential moving
average with coefficient $\lambda$ on the trajectory of the detrended
log-return $\xi_t$. The process $\v\eta_t = (y_t; q_t)$ has dynamics
\begin{align*}
  \v\eta_t &= -\tilde{\vv{A}}(\v\eta_t - \v\eta_{\infty})dt
  + \pmx{ 1 \\ 2y_t} \sigma_tdW_t,
  &
  \tilde{\vv{A}} &= \pmx{ \lambda & 0 \\ -2\beta & 2\lambda - \gamma },
  &
  \v\eta_{\infty} &= \pmx{ 0 \\ q_{\infty}},
\end{align*}
and where $q_\infty = \alpha/(2\lambda(1-\kappa))$,
$\kappa = \gamma/(2\lambda)$. The mean reversion level for
$\sigma_t^2$ is given by
$\sigma_\infty^2 = \alpha/(1-\kappa) = 2\alpha\lambda/(2\lambda -
\gamma)$. Moreover, $\vv{A}_{33}$ and $\vv{A}_{44}$ are scalar and
their values are given by
\begin{align*}
  \vv{A}_{33} &= 3\lambda - 2\gamma,
  &
  \vv{A}_{44} &= 4\lambda - 6\gamma.
\end{align*}
It follows that the stability condition becomes
\begin{align*}
  \gamma < 2\lambda/3.
\end{align*}
Note that, the constants $\kappa$ and $\tilde\kappa$ defined in
\cref{eq:mv:qinf} and \cref{eq:mv:tkappa} are given by
$\kappa = \gamma/(2\lambda)$ and $\tilde\kappa = \gamma/\lambda$.

Thus, the variance process has two roots, with values $\lambda$ and
$2\lambda-\gamma$. When the stability condition, the additional root
belongs to the range $[4\lambda/3, 2\lambda]$.

Note that, for the stationary solution, $y_\infty = \E[y_t] = 0$ but
$\sigma_{\infty}^2 = \E[\sigma_t^2]$ is different from the value of
$\sigma_t^2$ when $y_t=0$. Moreover, if $\beta\neq 0$, the equilibrium
value for $y_t$ does not correspond to a minimum volatility level, see
\cref{tab:scalar:1}. Note that, the stationary variance linearly
depends on the coefficient $\alpha$ and it is also always larger:
$\sigma^2_\infty = \alpha(1-\kappa) > \alpha$, since $\kappa >0$.
\begin{table}[htbp]
  \footnotesize
  \caption{Specific values for $y_t$ and $\sigma_t$}
  \label{tab:scalar:1}

  \begin{center}
  \begin{tabular}{lccc}
    \toprule
    Description of state & $y_t$ & $q_t$ & $\sigma_t^2$ \\
    \midrule
    $y_t$ at mean-reversion level & 0 & 0 & $\alpha$
    \\
    $y_t$ at stationary distribution & 0 & $q_\infty$ & $\sigma_\infty^2$
    \\
    $y_t$ at minimum spot-variance & $-\beta/\gamma$ & 0 & $\alpha - \beta^2/\gamma$
    \\
    \bottomrule
  \end{tabular}
  \end{center}

\end{table}

Some computations lead to the following expressions for
$m^{(3)}_\infty$ and $m^{(4)}_\infty$, respectively, the third and
fourth order moments of $\v{y}_t$ under the weak-stationary
distribution,
\begin{align*}
  m^{(3)}_\infty  &= \frac{2\beta}{\lambda-\gamma}q_\infty,
  &
  m^{(4)}_\infty  &= 3
  \frac{2\lambda-\gamma}{2\lambda - 3\gamma}
  \Big(
  1 + 4 \frac{\beta^2}{\alpha(\lambda-\gamma)}
  \Big)
  q_\infty^2.
\end{align*}
Clearly, the parameter $\beta$ directly controls the asymmetry of the
distribution of $y_t$. 

\begin{figure}[htb]
  \centering

  \includegraphics[scale=0.58]{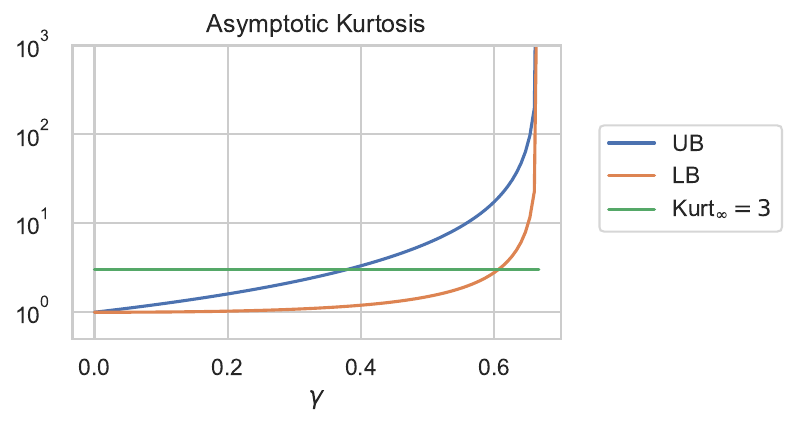}
  
  \label{fig:scalar:kurt}
  
  \caption{The upper- (UB) and lower-bound (LB) of the asymptotic
    Kurtosis given in \cref{eq:scalar:kurtBounds} as function of
    $\gamma$, when $\lambda=1$. The green horizontal line represents
    the mesokurtic level.}
\end{figure}

It can be shown, that under stationarity, the returns Kurtosis defined
in \cref{eq:mv:kurtosis} is given by
\begin{align}
  \text{Kurt}_\infty
  &= 
  \frac{2\lambda-\gamma}{2\lambda - 3\gamma}
  \Big(
  \frac{\lambda- \gamma}{\lambda}
  + 
  \frac{2\lambda -\gamma}{\lambda-\gamma}
  \frac{\beta^2}{\lambda\alpha}
  \Big),
  \label{eq:scalar:kurtInf}
\end{align}
W.r.t. $\beta$, that kurtosis is minimal when $\beta=0$, maxmimal when
$\beta^2 = \gamma\alpha$. Thus, the followning bound hold:
\begin{align*}
  \frac{2\lambda-\gamma}{2\lambda - 3\gamma}
  \frac{\lambda- \gamma}{\lambda}
  \leq
  \text{Kurt}_\infty
  &\leq
  \frac{\lambda}{\lambda-\gamma}
  \frac{2\lambda-\gamma}{2\lambda - 3\gamma}.
\end{align*}
Note also that $\text{Kurt}_\infty$ has the following homogeneity
property
\begin{align}\label{eq:scalar:kurtBounds}
  \text{Kurt}_\infty
  &= 
  \frac{2-\tilde\gamma}{2 - 3\tilde\gamma}
  \Big(
  1- \tilde\gamma
  + 
  \frac{2 -\tilde\gamma}{1-\tilde\gamma}
  \frac{\tilde\beta^2}{\tilde\alpha}
  \Big),
\end{align}
where $\tilde\alpha = \alpha/\lambda$, $\tilde\beta=\beta/\lambda$ and
$\tilde\gamma=\gamma/\lambda$. The upper- and lower-bound given in
\cref{eq:scalar:kurtBounds} are shown in \cref{fig:scalar:kurt}.
Notice that, the value of $\gamma/\lambda$ is the main driver for the
Kurtosis and that any possible value of the kurtosis can be obtained
by this model.

Resuming, the stationary distribution of the scalar model has the
following qualitative behaviour:
\begin{itemize}
\item %
  under stationarity, expectation does not corresponds to minimum
  volatility unless $\beta=0$;
\item %
  asymmetry of $y_t$ is directly proportional to $\beta$;
\item %
  return's kurtosis is an increasing function of $\gamma$ and of
  $\beta^2/\alpha$.
\end{itemize}

\begin{proposition}[Stationary distribution of the scalar QHR model]
  When $\Delta = \alpha\gamma - \beta^2 > 0$ and
  $\gamma < 2\lambda/3$, the SDE \cref{eq:scalar:dy} has a stationary
  solution given by a Pearson type IV distribution. More precisely,
  the stationary distribution $y_\infty$ has probability density
  function (PDF) given by
  \begin{align*}
    p_{y_{\infty}}(y) = \Prob[y_\infty \in dy] 
    &=
    C(\alpha + 2\beta y + \gamma y^2)^{-\frac\lambda\gamma-1}
    \exp\!\Big(
    \frac{2\lambda\beta/\gamma }{ \sqrt{\Delta} }
    \arctan\big(\frac{\beta + \gamma y}{\sqrt{\Delta}} \big)
    \Big),
  \end{align*}
  where $C$ is a normalisation constant.

  In particular, when $\beta=0$, $y_{\infty}$ is proportional to a
  Student-t random variable with $2\lambda/\gamma + 1$ degrees of
  freedom:
  \begin{align*}
    \sqrt{\frac\alpha{2\lambda+\gamma}}
    y_\infty
    \sim \text{Student-t}(2\lambda/\gamma+1).
  \end{align*}

  \begin{proof}
    It is direct to verify that $p_{y_\infty}$ is positive and solves the
    Pearson equation
    \begin{align*}
      \frac{p'(y)}{p(y)}
      &= - 2
      \frac{
        \beta + (\lambda+ \gamma)y
      }{
        \alpha + 2\beta y + \gamma y^2
      }
    \end{align*}
    (see \cite{Pearson:1895,PearsonMathWorld}). The latter can be
    rewritten as
    \begin{align*}
      \frac12 \partial_y (\sigma^2(y) p(y)) + \lambda y p(y) &= 0.
    \end{align*}
    It follows that $p(y,t) = p_{y_\infty}(y)$ is a constant in time
    solution to the Fokker-Planck equation
    \cref{eq:scalar:dy}:
    \begin{align*}
      \frac12 \partial_{yy}(\sigma^2(y) p(y,t))
      + \partial_y (\lambda y p(y,t)) &= \partial_t p(y,t).
    \end{align*}
    Thus, $p_{y_\infty}$ is the density of a stationary distribution
    for the SDE \cref{eq:scalar:dy}.
  \end{proof}
\end{proposition}

\subsubsection{Numerical tests}

\begin{table}[htbp]
  \footnotesize
  \caption{Setups used for the experimental results.}
  \label{tab:scalar:models}

  \begin{center}  
    \input{qhr_table.tex}
  \end{center}
\end{table}
In the rest of this subsection a few examples are tested to show the
capabilities of the scalar model. The parameter setup of those models
are reported in \cref{tab:scalar:models}. All those models allow for a
stationary distribution, since in all the four cases
$2\lambda-3\gamma >0$. The first two of those models, namely $M_1$ and
$M_2$, are symmetrical. All the models have the same exponential
parameter $\lambda=6$, except for $M_2$ whose parameter is
$\lambda=4$. The non-symmetrical models, namely $M_3$ and $M_4$, have
the minimum spot volatility level of $5\%$ is reached when
$y_t = 0.06$. Roughly speaking, the market is most quiet when the
underlying price is 6\% above its exponential moving average. Those
two models, despite their lower minimum volatility level, 5\% vs 8\%,
have larger stationary variance and kurtosis than those of the first
two models.

\begin{figure}[htbp]
  \centering

  \includegraphics[scale=.58]{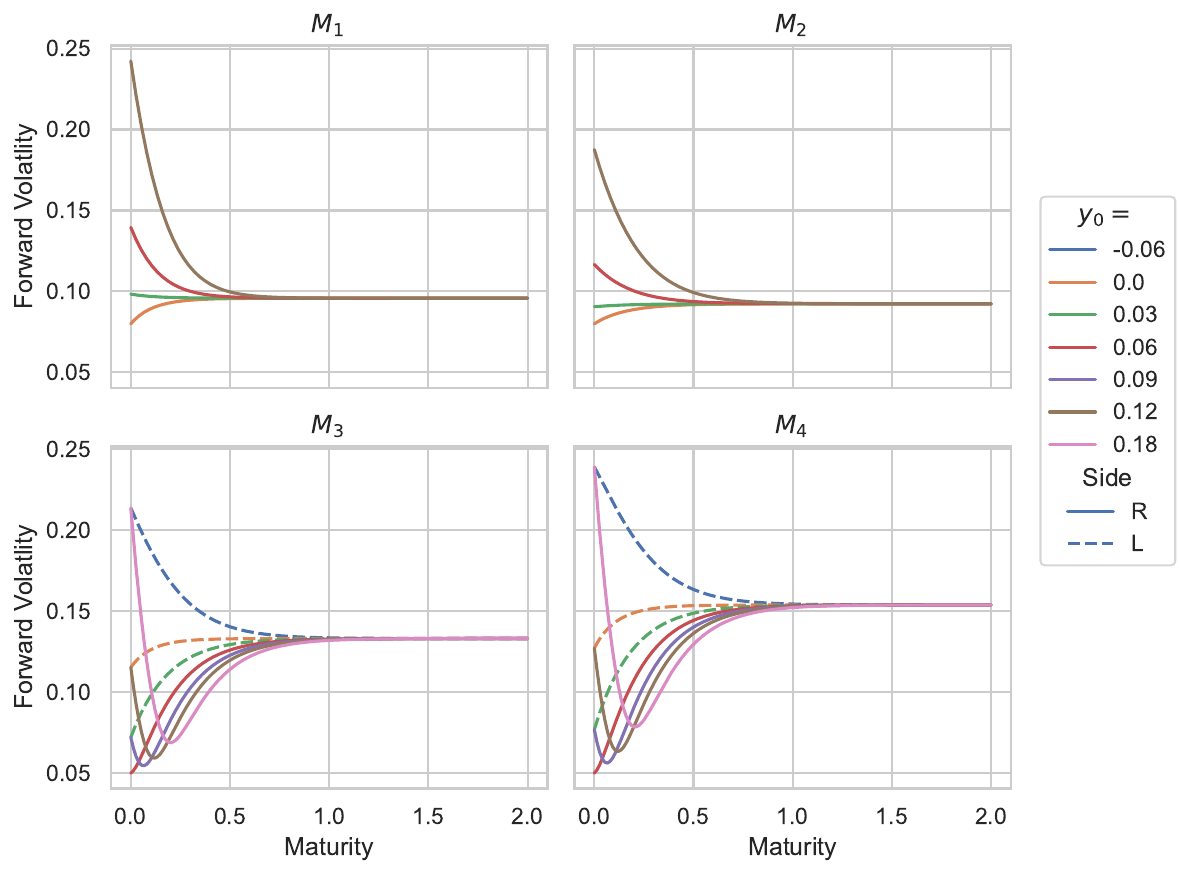}

  \caption{%
    The forward volatility curve $\sqrt{v_0(T)}$ under models
    $M_1, M_2, M_3$ and $M_4$ defined in \cref{tab:scalar:models}.
    Different colours represent different values for the initial
    condition $y_0$. Solid and dashed lines indicate, respectively
    $y_0 \geq -\beta/\gamma$ (Side = R) and $y_0< -\beta/\gamma$
    (Side = L). }
  \label{fig:scalar:fwdCurves}
\end{figure}

\Cref{fig:scalar:fwdCurves} shows the forward volatility curves
$T \to \sqrt{v_0(T)}$ of the four models for different values of the
initial condition $y_0$. Note that the asymmetric models show a richer
behaviour with non-necessarily monotone forward curves.

\begin{figure}[htbp]
  \centering

  \includegraphics[scale=.57]{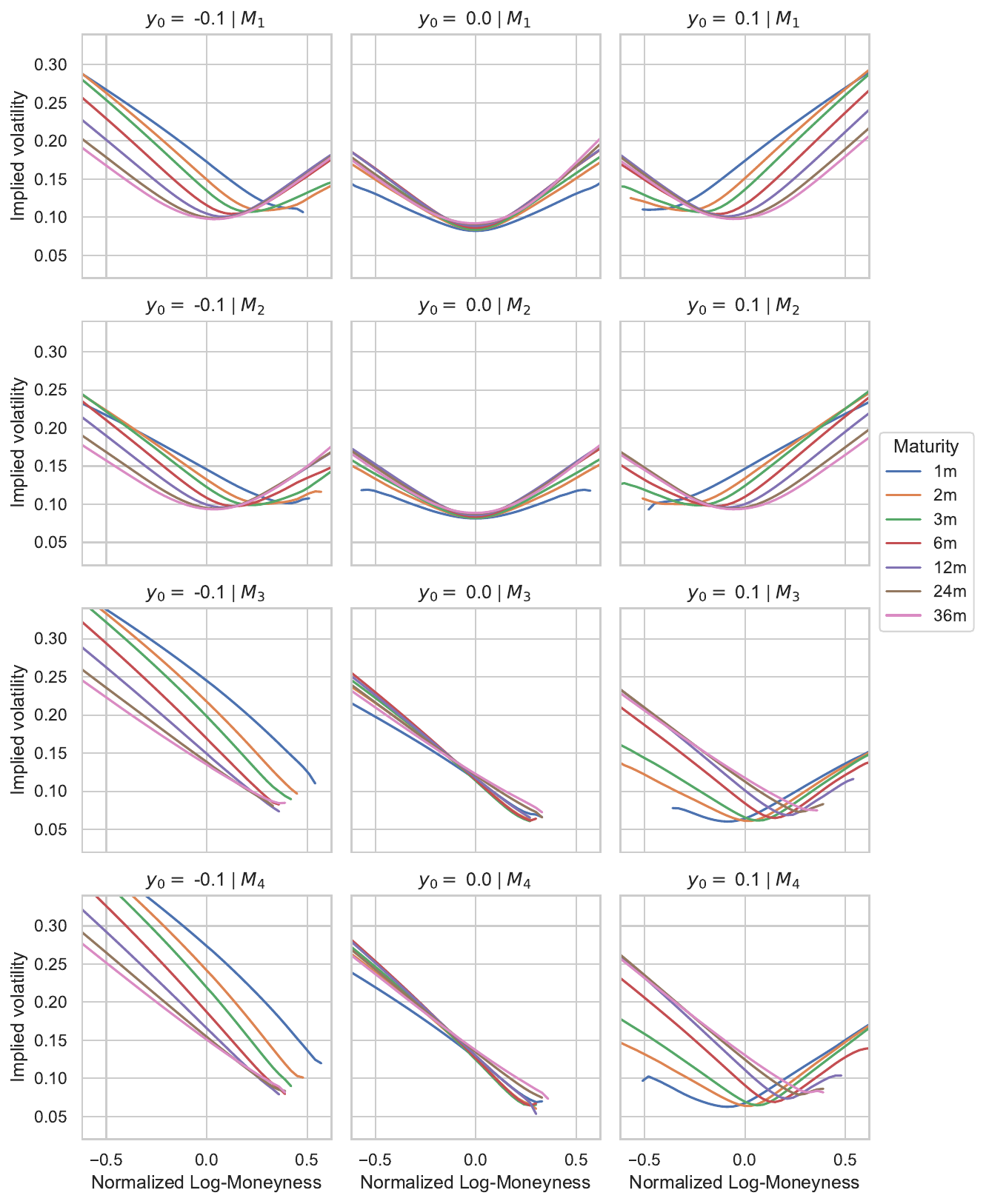}
  
  \caption{%
    Implied volatility smiles under models $M_1 - M_4$ (rows) in terms
    of the Log-Moneyness $\log(K) - x_0$ with $K$ is option exercise
    price. The price of the underlying is $e^{x_0} = 1$ and level of
    the offset is reported on top of each plot ($y_0=-.1$ on the left,
    $y_0=0$ at the center and $y_0=.1$ on the right). Implied
    volatilities refer to prices computed by means of an Euler-MC
    method with antithetic variables using 250 time-steps per year and
    $10^6$ trajectories. }
  \label{fig:scalar:smiles}
\end{figure}

\begin{figure}[htbp]
  \centering

  \includegraphics[scale=.58]{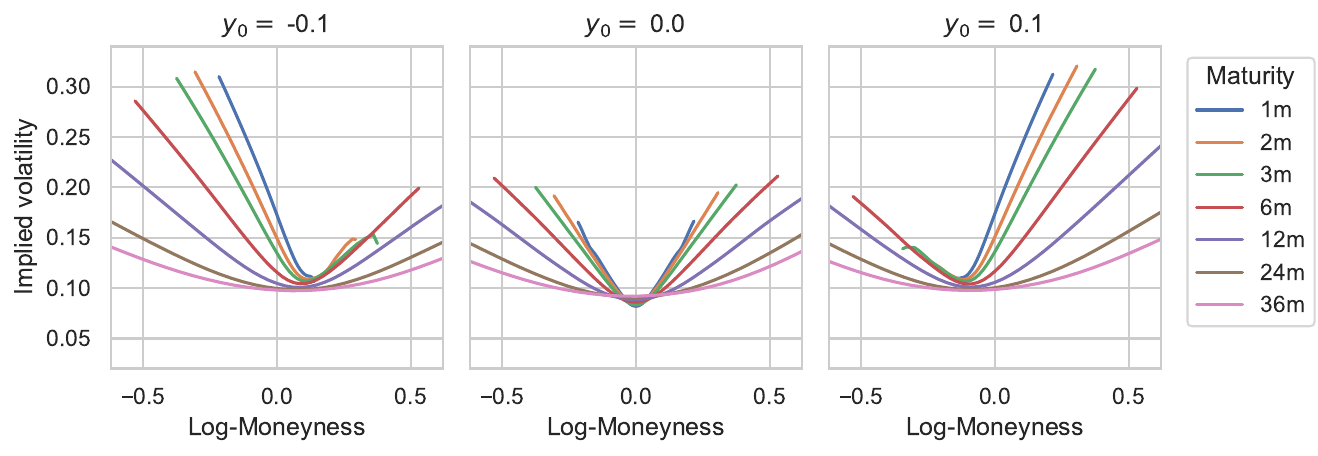}
  
  \caption{%
    Implied volatility smiles under models $M_1 - M_4$ (rows). The
    price of the underlying is $e^{x_0} = 1$ and level of the offset
    is reported on top of each plot. On the x-axis the ``Normalised
    Log-Moneyness'' $\log(K)/\sqrt{T}$, with $K$ being the strike
    price. Option prices computed by an Euler-MC method with
    antithetic variables using 250 time-steps per year and $10^6$
    trajectories. }
  \label{fig:scalar:smiles2}
\end{figure}

To evaluate the qualitative properties of the scalar QHR when pricing
European options, \cref{fig:scalar:smiles2} shows the implied
volatility smiles for models $M_1 - M_4$.
For each model (row), the smiles for different maturities $T$ (color)
and different initial conditions $\v{y}_0$ (column) are plotted in
terms of the ``Normalised Log-Moneyness'' $\log(K/S_0)/\sqrt{T}$.
That standardisation has the advantage of letting the smiles more
comparable across maturities. To support that statement, compare the
first row of \cref{fig:scalar:smiles2} with \cref{fig:scalar:smiles}
where the same smiles are plotted as function of the plain
log-moneyness $\log(K/S_0)$.
Recall that, by the homogeneity properties discussed in section
\ref{sec:homo}, the pricing function depends only on the $K/S_0$ and
not separately on the underlying price $S_0$ and strike price $K$. For
this reason all the prices have been computed for an underlying
$S_0=1$ ($x_0=0$). In all those experiments, prices are computed by
means of a an Euler Monte Carlo method with Antithetic Variables, a
time step of $1/250$ and $10^6$ scenario. For each model and initial
condition, all the prices are computed on the same set of
trajectories.

As expected, $\beta\neq 0$ leads to asymmetrical shapes and a larger
value of $\gamma$, as in models $M_1$ and $M_4$, corresponds to a
steeper smile shape. Moreover, in the asymmetric models $M_3$ and
$M_4$, when the offset value becomes substantially negative, a larger
$\gamma$ gives higher implied volatilities. Clearly, the smirks
obtained by the asymmetrical models $M_3$ and $M_4$ more realistically
replicate market stylised facts. However, even in those models, as the
offset becomes large and positive (here $y_0=0.1$) some smile looses
its monotonous characteristics. Recall that, for those models, the
spot volatility minimum is located at an offset $y_t = 0.06$.

\begin{figure}[htbp]
  \centering

  \includegraphics[scale=.58]{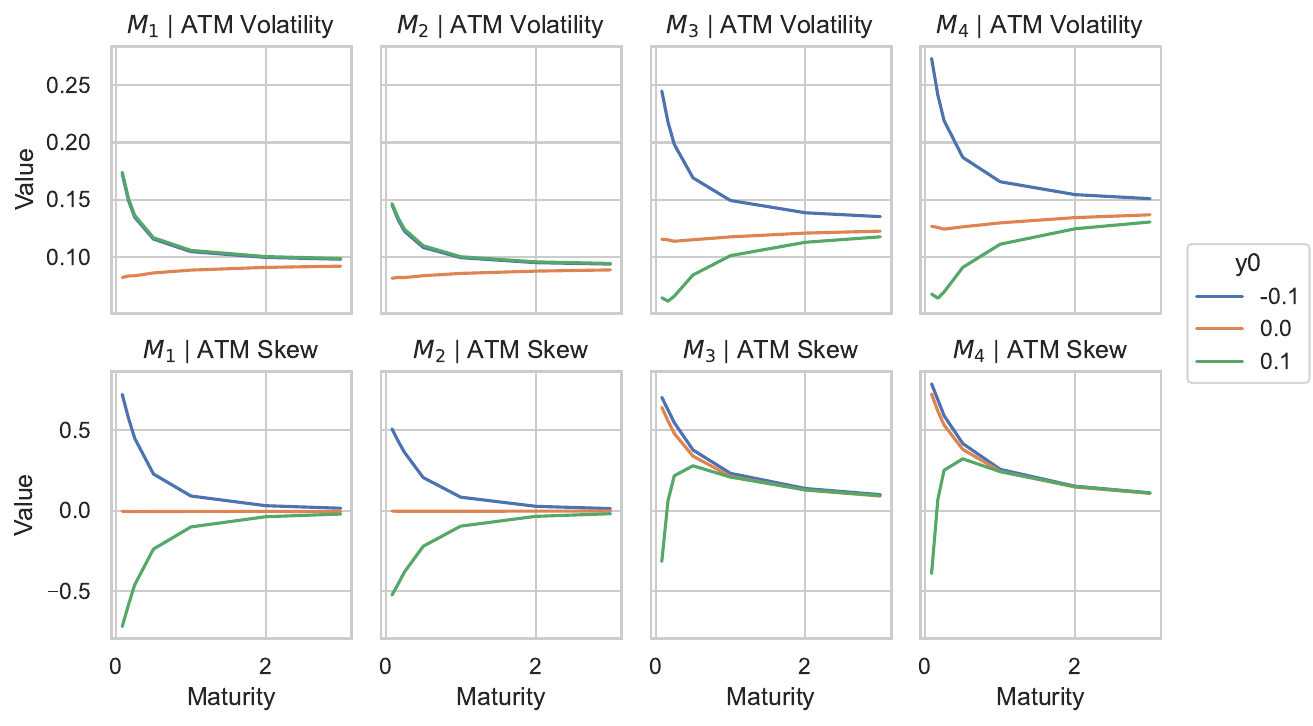}
  
  \caption{%
    Term structure of ATM-Volatility (top) and ATM-Skew
    (bottom) as defined in \cref{eq:scalar:ATMTerm}. }
  \label{fig:scalar:atmvol}
\end{figure}

Next, \cref{fig:scalar:atmvol} shows on the top the term structure of
the ATM implied volatility and its skew. More precisely, if
$\sigma_{impl}(\ell, T)$ denotes the implied volatility for
log-moneyness $\ell = \log(K/S_0)$ and maturity $T$, the
ATM-Volatility and ATM-Skew term structures are, respectively, the
functions
\begin{align}\label{eq:scalar:ATMTerm}
  \text{ATM-Volatility}\!: 
  T &\to \sigma_{impl}(0,T)
  &\text{and}&&
  \text{ATM-Skew}\!: 
  T &\to \frac{ d \sigma_{impl}(x,T) }{d\ell} \Big|_{\ell=0}.
\end{align}
In these tests, that derivative is approximated by finite differences.
The symmetry of models $M_1$ and $M_2$ does not deserve further
comments. The following remarks refers to the asymmetric models $M_3$
and $M_4$. Firstly, the ATM volatility qualitatively parallels the
forward volatility shapes seen in \cref{fig:scalar:fwdCurves}.
Next, in neutral ($y_0=0$) and ``bear'' ($y_0=-0.1$) market states the
Skews term structures are both monotonically decreasing and not very
different, qualitatively confirming stylised facts observed on market
skews. However, when the market had a ``bull'' trajectory, that
decreasing pattern is inverted for short maturities.

\subsection{Rank-one QHR model}

Consider a QHR model where the variance function depends on a single
offset: $\eta_t = \v{w}^T\v{y}_t$ where $\v{w} \in \Real^p_+$ is such
that $\v{w}^T\v{b} = 1$. In this case, the variance function has the
shape given in \cref{eq:mv:vol2} with $r=1$ and $\beta_0, \gamma_0$
scalar.
This model will be called the Rank-One QHR (R1QHR) model. 

As described in the introduction, the offset process $\tilde{y}_t$ can
be interpreted as a moving average using the filter $\phi$ defined in
\cref{eq:intro:tildey}. That is
\begin{align*}
  \tilde{y}_t &= \xi_t - \int_{\infty}^t \phi(t-s)\xi_s ds,
\end{align*}
because $\v{w}^T\v{b}= 1$. Moreover, the second part of \cref{h:1}
reduces to the constraints $\alpha > 0$ and
$\beta_0^2 \leq \alpha\gamma_0$.

Notice that, although $\sigma_t^2$ is function of the scalar process
$\eta_t$, the dimension of the SDE \cref{eq:mv:dy} cannot be reduced
without loosing its Markovian property. Indeed, the dynamics of
$\eta$, that is
\begin{align*}
  d\eta_t &= \v{w}^T\vv\Lambda \v{y}_tdt + \sigma_t dW_t,
\end{align*}
has a trend that does not depend on $\eta_t$ only.

In the distinct roots and single root special cases the constants
$\kappa$ and $\tilde\kappa$ defined in \cref{eq:mv:qinf} and
\cref{eq:mv:tkappa} reduce to
\begin{align*}
  &\text{Distinct Roots:}
  &
  \kappa &= \gamma_0 \sum_{i,j=1}^p \frac{1}{\lambda_i + \lambda_j}w_iw_j,
  &
  \tilde\kappa &=
  \gamma_0 \lambda_{\min}
  \big(\sum_i^p \frac{w_i}{\lambda_i}\big)^2,
  \\
  &\text{Single Root: }
  &
  \kappa &= 2\frac{\gamma_0}{\lambda_1}
  \sum_{ij}  \binom{i+j-2}{i-1}
  \frac{w_i}{2^i} \frac{w_j}{2^j}
  &
  \tilde\kappa &=
  \frac{\gamma_0}{\lambda_1}\big(\sum_{i=1}^pw_i\big)^2.
\end{align*}

In the case of distinct roots with $\v{w}>0$, the roots of
$\tilde{\vv{A}}$ of the rank-one model are all reals. Indeed, in that
case matrix $\vv{A}_{22} = \bar{\vv\Lambda} - \ones \v\gamma^T$ with
$\bar{\vv\Lambda}$ diagonal and $\v\gamma \geq 0$. It is not hard to
prove that, for that reason, all the eigenvalues of $\vv{A}_{22}$ are
real (see \cref{thm:DiagonalPerturbation} in \cref{sec:proofs}).

\subsubsection{Numerical Tests}

The above R1QHR model is here numerically tested on a set of models
with two offset processes ($p=2$). \Cref{tab:r1:models} shows the
setup used on these five models. The first four models, namely
$MM_1, \ldots MM_4$, have two distinct roots ($n=2$, $m=1$), while
model $MM_5$ have a single root ($n=1$) with double multiplicity
($m=2$). In that table the first six columns reports the parameters
used for each model, while the remaining ones some statistics on them.
In particular, $\mu_2, \mu_3$ and $\mu_4$ are the smallest real part
of the eigenvalues of $\vv{A}_{22}$, $\vv{A}_{33}$ and $\vv{A}_{44}$,
respectively. The successive column shows the value of $\kappa$. Note
that, even though $\kappa$ is not always smaller than $1/3$ all the
models are stable since $\mu_2, \mu_3$ and $\mu_4$ is positive. The
column $\v{y}_{\min}$ shows a value for $\v{y}$ that minimises the
variance function. Being not strictly convex, the variance minima
belong to a line passing through $\v{y}_{\min}$ orthogonal to $\v{w}$.
The successive columns show the minimum value for the spot volatility,
and the asymptotic (stationary) volatility and kurtosys.
\begin{table}[htbp]
  \footnotesize
  \caption{Setups used for the experimental results.}
  \label{tab:r1:models}

  \begin{center} 
    \input{r1qhr_table.tex}
  \end{center}
\end{table}

The setup is designed along the following lines. The comparison starts
with a symmetrical model, namely $MM_1$, that satisfies the stability
sufficient condition $\kappa < 1$. The model is based on a filter
which has two roots. The first one ($\lambda_1=1$) allows to intercept
medium-term effects and the second one ($\lambda_2=6$) to short-term
effects. The short-term offset has larger weight ($w_2=0.8$ vs
$w_1=0.2$) and thus the volatility is more sensible to fast movements
than to deviations from medium-term trends.

Looking at \cref{tab:r1:models}, this models does not show heavy
tails under the stationary regime (Kurt$_\infty < 3$). In model
$MM_2$, the variance coefficient $\gamma$ is chosen to that
$\kappa = 1/3$. Nevertheless, the model is still stable, as
$\mu_2, \mu_3$ and $\mu_4$ are positive. Note that, increase on the
convexity of the variance function does not change too much the
asymptotic volatility, but it has a large effect on the kurtosys.

Model $MM_3$ is obtained from $MM_2$ by increasing the coefficient
$\alpha$ and moving the minimum of the variance function. In that
model has the volatility can reach a smaller value (5\%) but with a
larger asymptotic volatility (14.70\%). Note that model $MM_3$ has a
very large asymptotic kurtosis.

Next, the fourth model, $MM_4$, is designed with aim to have the same
characteristics of model $MM_3$ but with a larger value for the short
memory root: $\lambda_2 = 12$. Indeed, the variance function have the
same minimum and similar asymptotic value. Differently from the
previous model, the condition $\kappa < 1/3$ is not satisfied neither
with equality, nevertheless the Kurtosis has a smaller value than in
model $MM_3$.

Finally, model $MM_5$ is analogous to model $MM_3$ but using a filter
with a root $\lambda_1=6$ with double multiplicity. There, to have a
positive filter that integrates to one, the weights are chosen so that
$w_1=1$ and $w_2 \in [0,1]$. 

\Cref{tab:r1:eigen} reports the filter roots $\lambda_1$ and
$\lambda_2$ (first two columns) and the eigenvalues of $\vv{A}_{22}$
for the five model. Notice that, the models with distinct roots have,
as expected, real eigenvalues and model $MM_5$ have two conjugate
complex roots. Looking at the last column, that eigenvalue is given by
the sum of the two filter roots, that is $\lambda_1+\lambda_2$, for
the distinct root case, and by the double of the root for the double
multiplicity case $2\lambda_1$.

\begin{table}[htbp]
  \footnotesize
  \caption{Eigenvalues of the matrices $\vv\Lambda$ (first two
    columns) and $\vv{A}_{22}$ (last four columns).}
  \label{tab:r1:eigen}
  
  \begin{center}  
  \begin{tabular}{lrrrllr}
    \toprule
    Model &  &  &  &  &  &  \\
    \midrule
    $MM_1$         & 1.00 &  6.00 & 1.89 & \, 6.20 & 10.91 & 7.00 \\
    $MM_2$, $MM_3$ & 1.00 &  6.00 & 1.83 & \, 5.79 & 10.58 & 7.00 \\ 
    $MM_4$         & 1.00 & 12.00 & 1.74 & 11.09 & 21.47 & 13.00 \\
    $MM_5$         & 6.00 &       & 7.15 & $12.93-0.96j$ & $12.93+0.96j$ & 12.00  \\
    \bottomrule
  \end{tabular}
  \end{center}
\end{table}

\begin{figure}[htbp]
  \centering

  \includegraphics[scale=.58]{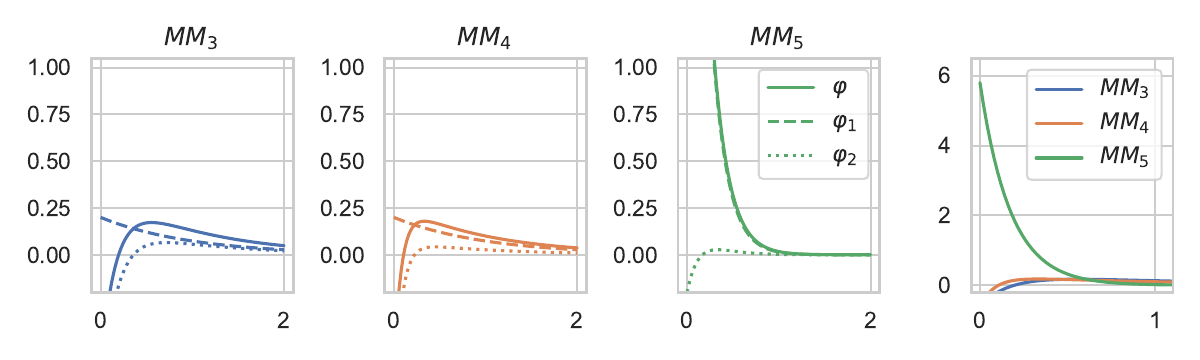}
  
  \caption{Plot of the filter $\phi$ and its components $\phi_1$ and
    $\phi_2$ for models $MM_3$, $MM_4$ and $MM_5$ as defined in
    \cref{eq:intro:tildey}. }
  \label{fig:r1:filters}
\end{figure}

The shape of the filters and of their components is shown in
\cref{fig:r1:filters} for models $MM_3, MM_4$ and $MM_5$. Those of
models $MM_1$ and $MM_2$ are not explicitly shown because identical to
that of $MM_3$. Note that, the filter $\phi_2$ of model $MM_3$ is
negative near the origin, but the whole filter $\phi$ is positive,
since $w_2 = 0.2 < w_1 = 1$. The qualitative properties of those
filters are the following. $MM_3$ and $MM_4$ have the similar decay
(long-memory) and $MM_5$ have a faster decay. On the short- and
medium-term side, $MM_3$ and $MM_5$ have similar weight, while $MM_4$
poses more weight on the short-term information and less on the
medium-term one.

\begin{figure}[htbp]
  \centering

  \includegraphics[scale=.58]{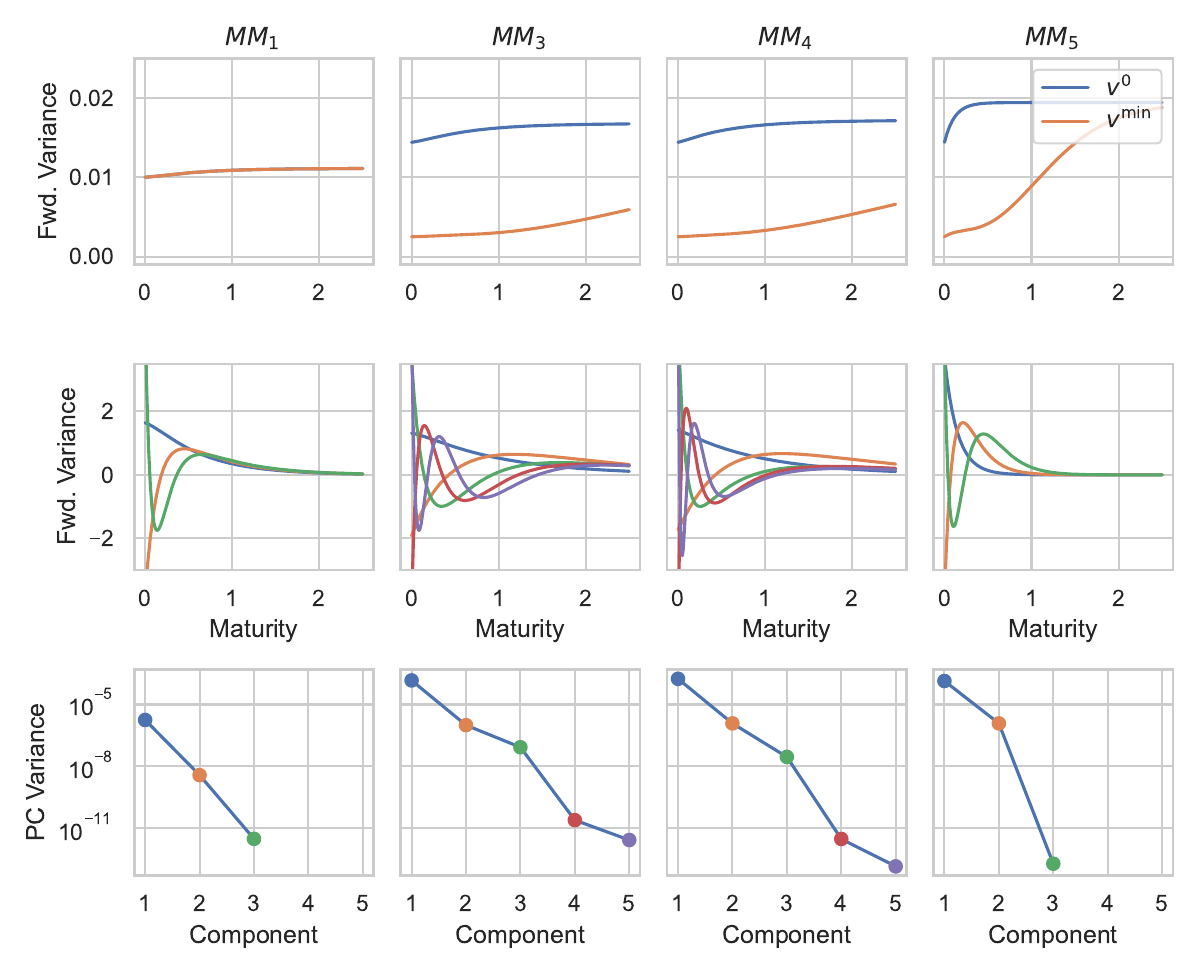}
  
  \caption{Term structure of instantaneous forward variance. Top:
    forward variance $v^0 = v_0(\cdot)$ when $y_0=0$ and at minimum.
    Middle: principal components (PC) of the forward variance curve.
    Bottom: variance of each PC. In the middle and bottom plots, the
    curve and the variance of each PC is shown using the same color
    (blue for the first PC, orange for the second one, green for
    the third one and so on).
  }
  \label{fig:r1:pca}
\end{figure}

Hereafter, only the performances of models $MM_3$, $MM_4$ and $MM_5$
are considered, since the symmetric models are not too much realistic.
\Cref{fig:r1:pca} shows the term structure of the forward
instantaneous variance. The plots in the top row show the curve $v^0$
and $v^{\min}$ defined in \cref{eq:mv:fwdVarDecomposition} and
\cref{eq:mv:fwdVarMin}. For model $MM_1$, the two curves are not
distinguishable because of the symmetry. Note that, since
$\lim_{s\to \infty}v_0(s) =\sigma_\infty^2$ for any
$\v{y}_t \in \Real^p$, both the curves converges asymptotically to
$\sigma_\infty^2$. At the scale of the plots, this behaviour is
clearly visible for models $MM_1$ and $MM_5$.

Another consequence of the symmetry is that since the forward curve
depends only on the three factors $y_{1t}^2$, $y_{2t}^2$ and
$y_{1t}y_{2t}$, and not directly on $y_{1t}$ and $y_{2t}$. That is,
$\v\psi^{(\v{y})}(s) = 0$. As a result, the PCA will have three
components with non-null variance as it is shown in the bottom left
plots of \cref{fig:r1:pca}.

Regarding models $MM_3$ and $MM_4$, the first principal component (PC)
have a quite fast decay. Then, those with smaller variance have a more
slow decay. Indeed, $\lambda_1=1$ was chosen with the purpose of
introducing a long-memory effect into the model. Next, compare the PCs
of model $MM_1$ and $MM_3$, that have the same roots $\lambda_1 = 1$
and $\lambda_2 = 6$, but different asymmetry and convexity. Those
differences results on a larger forward curve variance, as seen on the
PC variance's plots, and on slower decay of the PCs, as shown by the
third (green) and fourth (red) PCs of model $MM_3$.
On the contrary, all the PCs of $MM_5$, the model with a single root
of double multiplicity, have a quite fast decay. Also, the variance of
the last PCs is quite small (bottom right plot). Moreover, the last
two PCs of model $MM_3$, having a very small variance, should have a
very small effect on the forward variance.
Finally note that, the oscillation of the PC curves is an expected
consequence of their orthogonality.

\begin{figure}[htbp]
  \centering

  \includegraphics[scale=.58]{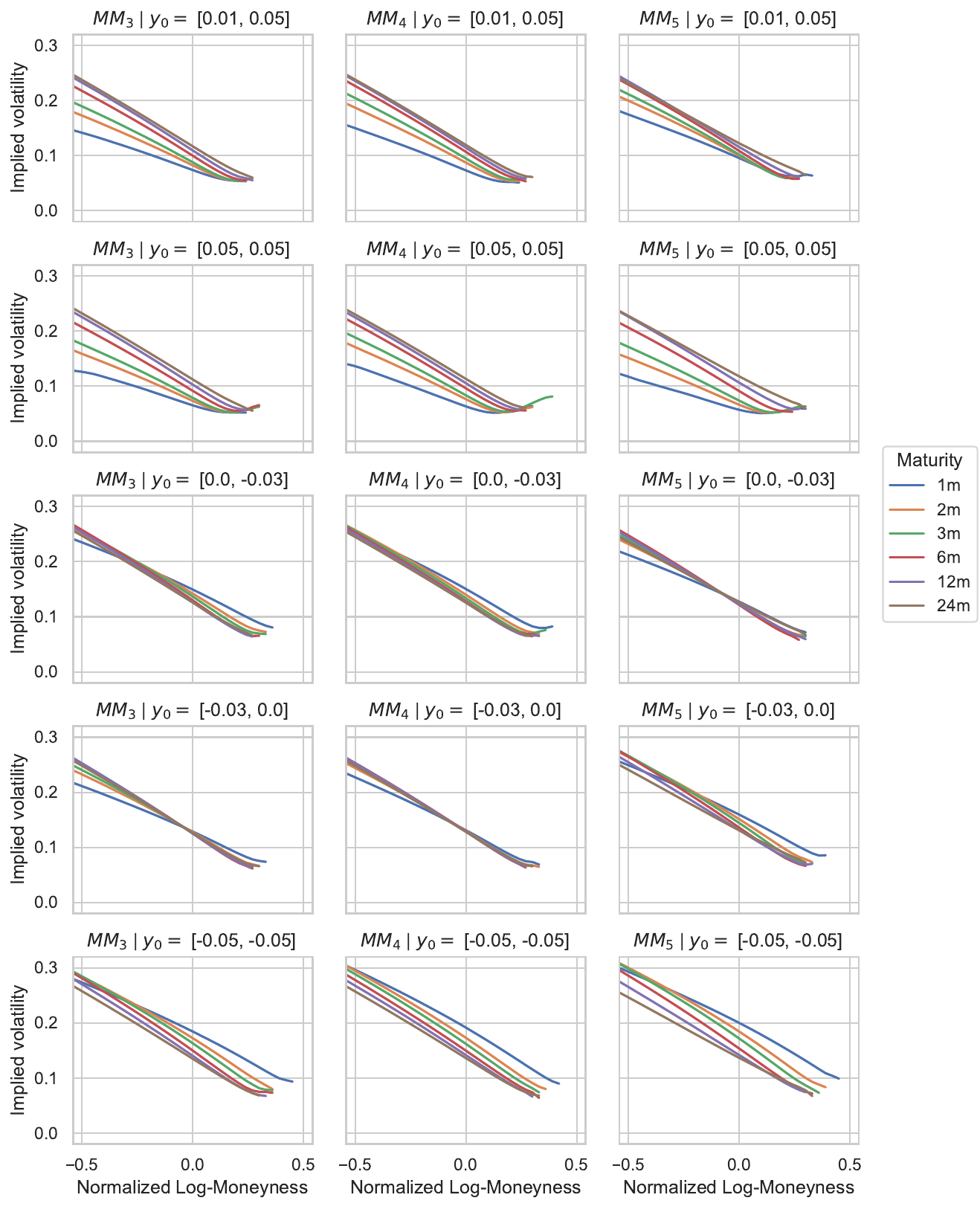}
  
  \caption{ Implied volatility smiles for models $MM_3$, $MM_4$ and
    $MM_5$ (columns), different initial conditions $\v{y}_0$ (rows)
    and maturities $T$ (color). The underlying price is $e^{x_0} = 1$
    and the implied volatilities are plotted against the ``normalised
    log-moneyness'' $\log(K)/\sqrt{T}$ where $K$ is the strike price.
    Option prices are computed by means of an Euler-MC method with
    antithetic variables, 250 time-steps per year and $10^6$
    trajectories. }
  \label{fig:r1:smiles}
\end{figure}

\Cref{fig:r1:smiles} shows the implied volatility smiles for the
asymmetric models $MM_3$, $MM_4$ and $MM_5$ for different initial
(market) conditions. Recall that for models $MM_3$ and $MM_4$ the two
elements of the initial condition $\v{y}_0$ corresponds to offsets
w.r.t. ``long'' term and short term memory averages. The first initial
condition $\v{y}_0 = (0.01, 0.05)$ corresponds to a market where, the
underlying is only slightly above the long term trend and consistently
above w.r.t. the short term one. For instance, the underlying had a
recent growth that allowed to return above its long-term average. The
second scenario, $\v{y}_0 = (0.05, 0.05)$, corresponds to a ``bull''
market where the underlying had a substantial growth w.r.t. both the
offsets. In the third scenario, $\v{y}_0 = (0, -0.03)$ the underlying
performances are not bad w.r.t. a long-term evaluation, but quite bad
for an investor who entered the market only recently. In the fourth
setup, $\v{y}_0 = (-0.03, 0)$ those two behaviour are inverted.
Finally, $\v{y}_0 = (-0.05,-0,05)$ represents a ``bear'' market
contingency where the underlying under-perform w.r.t. both the
averages.
Regarding model $MM_5$, the first offset is an exponential moving
average with memory parameter $\lambda=6$ as in \cref{eq:mv:y1}. As
discussed in Section \ref{sec:filters}, the second element of
$\v{y}_t$ is the offset between the exponential moving average and a
longer memory moving average whose weight is more concentrated at a
lag of 1/6 years, that is 2 months. At a different time scale, the
shape of these two filters is shown in the left plot of 
\cref{fig:mv:filt0}, as a blue and yellow curve.

Regarding the smiles shown in \cref{fig:r1:smiles}, the consequences
of bear market, $\v{y}_0 = (-0.05,-0.05)$, are an increase in all the
volatility smiles (last row). Comparing the different row of
\cref{fig:r1:smiles}, it seems that a variation on only one of the two
offsets does not have similar substantial consequences across maturity
and moneyness. Rather, that the effects of those changes are a
different skew or a later shift of the smile curve. For instance, in
the mixed market conditions, $\v{y}_0 = (0, -0.03)$ and
$\v{y}_0 = (-0,03, 0)$, the smiles of short-term and long-term
maturities crosses (see 3rd and 4th rows of \cref{fig:r1:smiles}).

\begin{figure}[htbp]
  \centering

  \includegraphics[scale=.53]{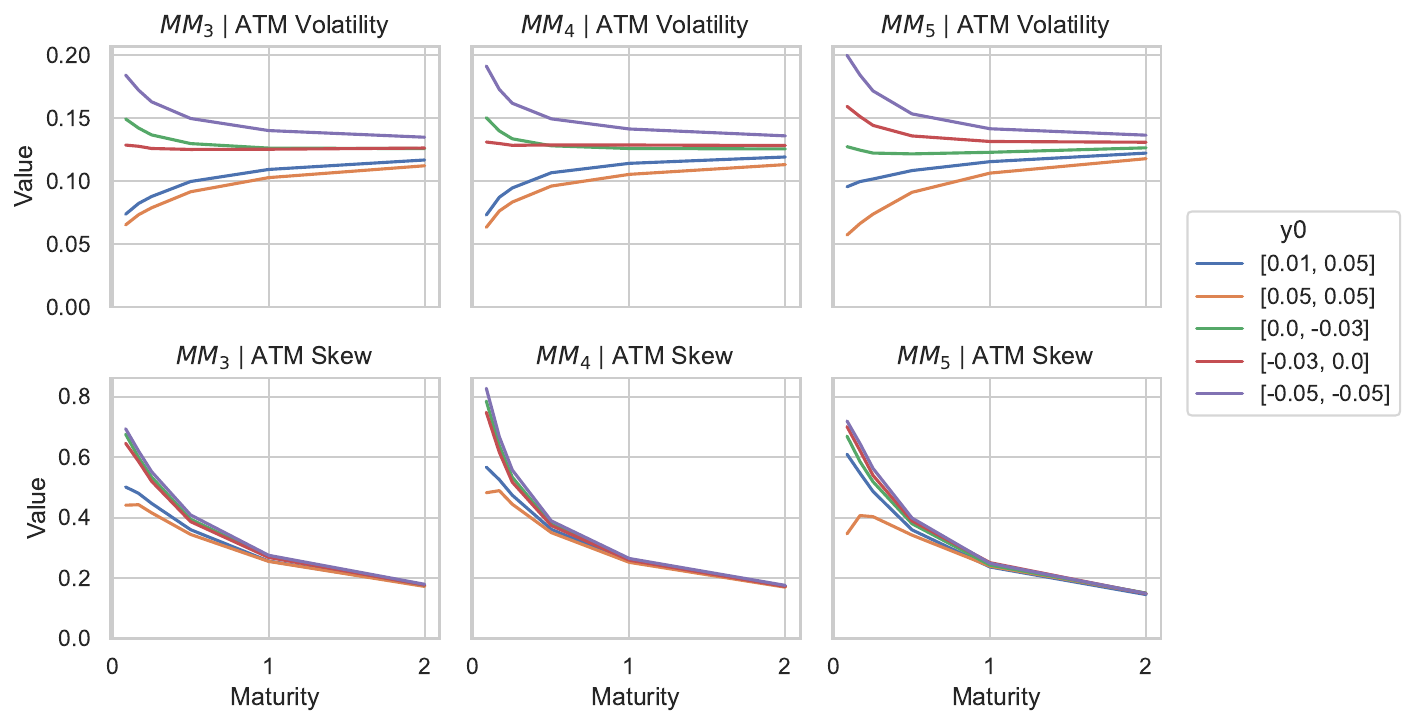}

  \caption{%
    Term structure of ATM Volatility (top) and ATM Skew for models
    $MM_3$, $MM_4$ and $MM_5$.
  }
  \label{fig:r1:atmvol}
\end{figure}

For the same set of R1QHR models and initial conditions, the
ATM-Volatility and ATM-Skew term structures defined in
\cref{eq:scalar:ATMTerm} are shown in \cref{fig:r1:atmvol}. As
expected, the larger value of $\lambda_2$ in model $MM_4$ do not
influence too much the long term behaviour of those term structures,
but only the short term one. Moreover, the ATM skew term-structure for
the model with double multiplicity seems to be quite different from
the other two. Qualitatively, the initial condition $\v{y}_0$ affects
both the curves homogeneously across models. The only exception is,
again, model $MM_5$ where the ATM curves of the two mixed market
conditions (blue and green) are inverted w.r.t. the other two models.

\section{Conclusions and further extensions}

This work develop the QHR model, a multifactor quadratic
generalisation of the HR model \cite{HobsonRogersR:98} with the
following properties:
\begin{itemize}
\item Markovianity;
\item autonomous and non-dimensional dynamics specification;
\item path dependent volatility;
\item price and volatility are diffusion processes;
\item weak-stationary volatility;
\item the variance and the squared increments processes have an ARMA
  autocorrelation structure;
\item allowing for lepto- and platy-Kurtic returns;
\item explicit expression for the forward variance;
\item flexible forward variance, ATM volatility and ATM skew curves.
\end{itemize}

The QHR model shares most of those properties with the COGARCH model
which requires a driving Levy process with discontinuous paths. Also
the quadratic specification for the variance is analogous to quadratic
rough-Heston model \cite{GatheralJusselinRosenbaum:2020}. Regarding
the absence of roughness in QHR volatility model, note that rough
models are often approximated by multifactor models analogous to the
QHR model \cite{AbiJaberElEuch:2019}. Moreover, to the knowledge of
the author, the presence of roughness in financial volatility
processes is still debated \cite{ContDas:2024}.

Although, the purpose of this paper is to present the model and study
the properties of the models, some numerical experiments are reported
to show the model capabilities. However, the market performances of
the QHR model deserve further research that will consist on the
calibration to forward and option prices or on an econometric fitting
on historical price trajectories.

Another direction to extend the model here presented is to consider a
fully second order model. More precisely, the QHR model is of second
order in the sense that the variance is a quadratic function of the
first order offset process $\v{y}_t$. However, a second order model
can be also obtained by introducing factor which is a moving average
of the squared offsets like the following
\begin{align*}
  \vv{Z}_t &= \int_{-\infty}^t
  (e^{-\frac12\vv\Lambda(t-s)})
  \v{y}_s\v{y}_s^T
  (e^{-\frac12\vv\Lambda(t-s)})^T
  ds.
\end{align*}
Note that, when $p=1$, the latter reduces to
$z_t = \int_{-\infty}^te^{-\lambda (t-s)}y_s^2ds$. Having defined
$\vv{Z}_t$, it could be included into the model by defining adding the
term $\tr(\vv\Delta \vv{Z}_t)$ to the variance specification
\cref{eq:mv:vol0}. Assuming $\vv\Delta \in \mathcal{S}_+^p$ is
sufficient to ensure the positivity of $\sigma_t^2$. It is worth
remarking that the original HR setup already considered the use of
offsets of some order $m>1$. However, all the work that originated
from it, including the present one, is based on first order offset
processes
\cite{FigaTalamancaGerra:2006,FoschiPascucci:2008,FoschiPascucci:2009,HalulliVargiolu:2008,RosestolatoVargioluVillani:2013}.

% ----------------------------------------------------------------
\bibliographystyle{amsplain}
\bibliography{references}
% ----------------------------------------------------------------

\clearpage

\appendix

\section{Auxiliary results and proofs}\label{sec:proofs}

\begin{proposition}\label{thm:DiagonalPerturbation}
  Let $\vv{D} = \diag(d_1,\ldots, d_n)$ and let $\v{x} \in \Real^n_+$.
  Then, all the eigenvalues of $\vv{D} - \ones\v{x}^T$ are real.
  \begin{proof}
    Without loss of generality assume that all the zero elements of
    $\v{x}$ are positioned at the top. Then, $\v{x}$, $\vv{D}$ and $\vv{D} -
    \ones\v{x}^T$ can be partitioned as
    \begin{align*}
      \v{x} &=  \pmx{ 0 \\ \v{x}_B},
      &
      \vv{D}  &= \pmx{ \vv{D}_A & 0 \\ 0 & \vv{D}_B},
      &
      \vv{D} - \ones \v{x}^T
      &= \pmx{
        \vv{D}_A & -\ones \v{x}_B^T \\
        0 & \vv{D}_B - \ones\v{x}_B^T
      },
    \end{align*}
    with $\v{x}_B>0$. We will prove that
    $\vv{M} = \vv{D}_B - \ones\v{x}_B^T$ is similar to a symmetric
    matrix. Define $\vv{X}_B = \diag(\v{x}_B)$. Then, $\vv{M}$ is
    similar to the matrix
    $\vv{X}_B^{\frac12}\vv{M}\vv{X}_B^{-\frac12} = \vv{D}_B -
    \vv{X}_B^{\frac12}\ones \ones^T\vv{X}^{\frac12}$, which is
    symmetric.
  \end{proof}
\end{proposition}

\begin{proof}[Proof of \cref{thm:mv:momentODE}]
  In the following the dynamics of $\v{y}^{\otimes k}$,
  $2 \leq k\leq 4$, is obtained form \cref{eq:mv:dy} by applying
  It\^o's Lemma. By \cref{thm:mv:standardHyp} also
  $\v{y}_t^{\otimes k}$ are non-exploding in finite time and also
  It\^o can be safely applied.
  
  \begin{itemize}
  \item $k=2$:

    Applying It\^o to the SDE \cref{eq:mv:dy} and the function
    $\v{y}_t^{\otimes 2}$ gives
    \begin{align*}
      d\v{y}_t^{\otimes 2}
      &=
      (-\vv\Lambda \v{y}_tdt + \v{b}\sigma_tdW_t) \otimes \v{y}_t
      + \v{y}_t \otimes (-\vv\Lambda \v{y}_tdt + \v{b}\sigma_tdW_t)
      + (\v{b}\sigma_t) \otimes (\v{b}\sigma_t) dt
      \\
      &=
      \big(
      (\v{b}\otimes \v{b})\sigma_t^2 
      - (\vv\Lambda \otimes \eye + \eye \otimes \vv\Lambda)\v{y}_t^{\otimes 2}
      \big) dt
      +
      (\v{b}\otimes \v{y}_t + \v{y}_t \otimes \v{b})\sigma_t dW_t
      \\
      &=
      \big(
      \alpha\bar{\v{b}}
      + 
      \vv{B}^{(2)}(2\v\beta^T\v{y}_t + \v\gamma^T\v{y}_t^{\otimes 2})
      - \vv\Lambda^{(2)}\v{y}_t^{\otimes 2}
      \big)
      dt
      +
      \vv{C}^{(2)}\v{y}_t \sigma_t dW_t,
    \end{align*}
    since, from \cref{eq:mv:LBCk},
    $\vv\Lambda^{(2)} = \eye_p \otimes \vv\Lambda + \vv\Lambda \otimes
    \eye_p$, $\vv{B}^{(2)} = \bar{\v{b}} = \v{b}\otimes \v{b}$,
    and $\vv{C}^{(2)} = \v{b} \otimes \eye_p + \eye_p \otimes \v{b}$.
    Then, using \cref{eq:mv:Ak},
    \begin{align}\label{eq:mv:dy2}
      d\v{y}_t^{\otimes 2}
      &= 
      (\alpha \bar{\v{b}}
      - \vv{A}_{12}\v{y}_t
      - \vv{A}_{22}\v{y}_t^{\otimes 2})dt
      + \vv{C}^{(2)}\v{y}_t \sigma_t dW_t.
    \end{align}
  \item $k=3$.
    
    From \cref{eq:mv:dy} and \cref{eq:mv:dy2} and applying It\^o,
    \begin{align*}
      d\v{y}_t^{\otimes 3}
      &= d(\v{y}_t \otimes \v{y}_t^{\otimes 2})
      \\
      &=
      \big(
      \alpha \v{y}_t \otimes \bar{\v{b}}
      - \v{y}_t \otimes (\vv{A}_{12}\v{y}_t)
      - \v{y}_t \otimes (\vv{A}_{22}\v{y}_t^{\otimes 2})
      \big)dt
      \\
      &\qquad
      +
      \big(\v{y}_t \otimes (\vv{C}^{(2)}\v{y}_t)\big)
      \sigma_t dW_t
      \\
      &\qquad      
      - (\vv\Lambda \v{y}_t) \otimes \v{y}_t^{\otimes 2}dt
      + \v{b}\otimes \v{y}_t^{\otimes 2} \sigma_t dW_t
      + \v{b} \otimes (\vv{C}^{(2)}\v{y}_t) \sigma_t^2 dt
      \\
      &=
      \big(
      \alpha (\eye_p\otimes \bar{\v{b}})\v{y}_t
      - (\eye_p \otimes \vv{A}_{12})\v{y}_t^{\otimes 2}
      - (\eye_p \otimes \vv{A}_{22})\v{y}_t^{\otimes 3}
      - (\vv\Lambda \otimes \eye_p^2)\v{y}_t^{\otimes 3}
      \big)dt
      \\
      &\qquad      
      + \v{b} \otimes (\vv{C}^{(2)}\v{y}_t) \sigma_t^2 dt      
      \\
      &\qquad      
      + (\v{b}\otimes\eye_{p^2})\v{y}_t^{\otimes 2} \sigma_t dW_t
      + (\eye_p \otimes \vv{C}^{(2)})\v{y}^{\otimes 2}
      \sigma_t dW_t
      \\
      &=
      \big(
      \alpha (\eye_p\otimes \bar{\v{b}})\v{y}_t
      - (\eye_p \otimes \vv{A}_{12})\v{y}_t^{\otimes 2}
      - (\eye_p \otimes \vv{A}_{22})\v{y}_t^{\otimes 3}
      - (\vv\Lambda \otimes \eye_p^2)\v{y}_t^{\otimes 3}
      \big)dt
      \\
      &\qquad
      +
      \big( 
      \alpha (\v{b} \otimes \vv{C}^{(2)})\v{y}_t
      + 2(\v{b} \otimes \vv{C}^{(2)})\v{y}_t \v\beta^T\v{y}_t
      + (\v{b} \otimes \vv{C}^{(2)})\v{y}_t \v\gamma^T\v{y}_t^{\otimes 2}
      \big)dt
      \\
      &\qquad      
      + \vv{C}^{(3)}\v{y}^{\otimes 2}\sigma_t dW_t.
    \end{align*}
    In the trend of the latter, the coefficients of $\v{y}_t$ is given by
    \begin{align*}
      \alpha (\eye_p\otimes \bar{\v{b}})
      +  \alpha (\v{b} \otimes \vv{C}^{(2)})
      = \alpha \vv{B}^{(3)} = - \vv{A}_{13},  
    \end{align*}
    from \cref{eq:mv:Ak}, \cref{eq:mv:Bk} and since $\bar{\v{b}} =
    \vv{B}^{(2)}$.
    Next, the term involving $\v{y}_t^{\otimes 2}$ is given by
    \begin{align*}
      \Big(
      - (\eye_p \otimes \vv{A}_{12})\v{y}_t^{\otimes 2}
      +& 2(\v{b} \otimes \vv{C}^{(2)})\v{y}_t \v\beta^T\v{y}_t
      \Big) dt
      \\
      &= 
      2\Big(
      (\eye_p \otimes \vv{B}^{(2)}\v\beta^T)\v{y}_t^{\otimes 2}
      + (\v{b} \otimes \vv{C}^{(2)})(\eye_p \otimes \v\beta^T)\v{y}_t^{\otimes 2}
      \Big) dt
      \\
      &= 
      2\Big(
      \eye_p \otimes \vv{B}^{(2)} + \v{b} \otimes \vv{C}^{(2)}
      \Big)
      (\eye_p \otimes \v\beta^T)\v{y}_t^{\otimes 2}
      dt
      \\
      &= 2\vv{B}^{(3)}(\eye \otimes \v\beta^T) \v{y}_t^{\otimes 2}dt
      \\
      &=  2(\vv{B}^{(3)} \otimes \v\beta^T) \v{y}_t^{\otimes 2}dt
      = - \vv{A}_{23}\v{y}_t^{\otimes 2}dt,
    \end{align*}
    from \cref{eq:mv:Ak} and \cref{eq:mv:Bk}.
    Analogously, the coefficient of $\v{y}_t^{\otimes 3}$ can be
    rewritten as follows
    \begin{align*}
      - \Big(
      \eye_p \otimes \vv{A}_{22}
      + \vv\Lambda\otimes \eye_{p^2}
      -& (\v{b}\otimes \vv{C}^{(2)})\otimes \v\gamma^T
      \Big)
      \\
      &=
      - \vv\Lambda^{(3)}
      + (\eye_p\otimes \vv{B}^{(2)}) \otimes \v\gamma^T
      + (\v{b}\otimes \vv{C}^{(2)}) \otimes \v\gamma^T
      \\
      &= - \vv\Lambda^{(3)} + \vv{B}^{(3)}\otimes \v\gamma^T
      = -\vv{A}_{33},
    \end{align*}
    since
    $(\v{b}\otimes \vv{C}^{(2)})\v{y}_t \v\gamma^T\v{y}_t^{\otimes 2}
    = ((\v{b}\otimes \vv{C}^{(2)})\otimes \v\gamma^T)\v{y}_t^{\otimes
      3}$.
    Then,
    \begin{align}\label{eq:mv:dy3}
      d\v{y}_t^{\otimes 3}
      &= - \Big(
      \vv{A}_{13}\v{y}_t + \vv{A}_{23}\v{y}_t^{\otimes 2} + \vv{A}_{33}\v{y}_t^{\otimes 3}
      \Big)dt
      + \vv{C}^{(3)}\v{y}_t^{\otimes 2}\sigma_t dW_t.
    \end{align}
  \item $k=4$.

    Analogously to the step $k=3$, applying It\^o's Lemma,
    rearranging the terms and exploiting the recursions
    \cref{eq:mv:Ak} and \cref{eq:mv:LBCk} gives
    \begin{align}\label{eq:mv:dy4}
      d\v{y}_t^{\otimes 4}
      = -\Big(
      \vv{A}_{24}\v{y}_t^{\otimes 2}
      + \vv{A}_{34}\v{y}_t^{\otimes 3}
      + \vv{A}_{44}\v{y}_t^{\otimes 4}
      \Big) dt
      + \vv{C}^{(4)}\v{y}_t^{\otimes 3}\sigma_t dW_t.
    \end{align}
  \end{itemize}
  Finally, the ODE \cref{eq:mv:muODE} is obtained from
  \cref{eq:mv:dy}, \cref{eq:mv:dy2}, \cref{eq:mv:dy3} and
  \cref{eq:mv:dy4} by taking the conditional expectations of
  $\v{y}_t^{\otimes k}$, $1\leq k \leq 4$.
\end{proof}

The proof of  \cref{thm:mv:stability} reported below will use the
following result.
\begin{proposition}\label{thm:mv:LInvBBound1}
  If  \cref{h:1,h:2} hold, then
  \begin{align}\label{eq:mv:LInvBBound1}
    \vv\Lambda^{-1}\v{b} > \lambda_{\min}^{-1}\v{b} \geq 0.
  \end{align}

  \begin{proof}
    Since, $\lambda_{\min}>0$ by \cref{h:1}, \cref{eq:mv:LInvBBound1}
    follows trivially from the structure of $\vv\Lambda$ and $\v{b}$
    in \cref{h:2} and from the fact that
    $\vv{D}_i^{-1}\v{b}_i= \ones \geq \v{b}_i$, $i=1,\ldots,m$.
  \end{proof}
\end{proposition}
\medskip

\begin{proof}[Proof of \cref{thm:mv:stability}]
  
  In the following we will use a vector $\v{z}$ chosen as follows:
  \begin{align*}
    \v{z} = \vv\Lambda^{-1}\v{b} + \v\xi,
  \end{align*}
  where $\v\xi$ is such that
  \begin{align}\label{eq:mv:AkPosXi}
    \v\xi &> 0,
    &
    \vv\Lambda \v\xi &> 0,
    &
    \v\xi^T \vv\Gamma \v\xi + \v\xi^T\vv\Gamma \vv\Lambda^{-1}\v{b}
    &< \frac{2-3\tilde\kappa}{3\lambda_{\min}}.
  \end{align}
  Since $\vv\Lambda$ is an $M$-matrix, it is semi-positive and, thus,
  there exists a vector satisfying the first two properties in
  \cref{eq:mv:AkPosXi}. Moreover, since the RHS of the third
  inequality in \cref{eq:mv:AkPosXi} is positive, such vector $\v\xi$
  can be rescaled to satisfy all the three inequalities. It follows
  that $\v{z}$ satisfies the following inequalities 
  \begin{align}\label{eq:mv:AkPosZ1}
    \vv\Lambda \v{z} &> \v{b},
    &
    \v{z} &> \vv\Lambda^{-1}\v{b},
    &\text{and}&&
    \v\gamma^T(\v{z} \otimes \v{z}) &< \frac2{3\lambda_{\min}}.
  \end{align}
  Moreover, note that under \cref{h:2}, by \cref{thm:mv:LInvBBound1},
  \begin{align}\label{eq:mv:AkPosZ2}
    \v{z} > \lambda_{\min}^{-1}\v{b} \geq 0.
  \end{align}
  
  The proof consists in showing that $\vv{A}_{22}$, $\vv{A}_{33}$ and
  $\vv{A}_{44}$ are $M$-matrices, and, thus, all their eigenvalues
  have positive real part \cite{Plemmons:1977}. Since
  $\v\gamma \geq 0$, the matrices $\vv{A}_{22}$, $\vv{A}_{33}$ and
  $\vv{A}_{44}$ are $Z$-matrices. That is, matrices with non-positive
  off-diagonal elements. To show that those are also $M$-matrices, it
  will be shown that they are semi-positive. That is,
  $\vv{A}_{kk}\v{x}_k > 0$ for some $\v{x}_k >0$ \cite{Plemmons:1977}.
  In particular, that property holds for
  $\v{x}_k = \v{z}^{\otimes k}$ with $k=2,3,4$. 
  \begin{itemize}
  \item $k=2$.
    The matrix product $\vv{A}_{22}\v{x}_2$ is given by
    \begin{align*}
      \vv{A}_{22}\v{x}_2
      &=
      (\vv\Lambda \otimes \eye + \eye \otimes \vv\Lambda -
      \bar{\v{b}}\v\gamma^T)
      (\v{z} \otimes \v{z})
      \\
      &>
      \v{b}\otimes \vv\Lambda^{-1}\v{b} + \vv\Lambda^{-1}\v{b}\otimes\v{b}
      -
      \frac2{3\lambda_{\min}} \v{b} \otimes \v{b}
      && \text{by \cref{eq:mv:AkPosZ1}}
      \\
      &\geq
      \frac4{3\lambda_{\min}}
      (\v{b}\otimes \v{b}),
      && \text{by \cref{eq:mv:AkPosZ2}}
    \end{align*}
    and thus $\vv{A}_{22}\v{x}_2>0$.
  \item $k=3$.
    Firstly note that
    $\vv{B}^{(3)} = 
    \v{b} \otimes \v{b} \otimes \eye
    +\v{b} \otimes \eye \otimes \v{b}
    + \eye \otimes \v{b} \otimes \v{b}$
    and, thus
    \begin{align*}
      \vv{B}^{(3)}(\eye \otimes \v\gamma^T)\v{z}^{\otimes 3}
      &=
      \v\gamma^T(\v{z}^{\otimes 2})
      ( \v{b} \otimes \v{b} \otimes \v{z}
      + \v{b} \otimes \v{z} \otimes \v{b}
      + \v{z} \otimes \v{b} \otimes \v{b} )
      \\
      &\leq
      \frac2{3\lambda_{\min}}
      ( \v{b} \otimes \v{b} \otimes \v{z}
      + \v{b} \otimes \v{z} \otimes \v{b}
      + \v{z} \otimes \v{b} \otimes \v{b} )
      && \text{by \cref{eq:mv:AkPosZ1}}
      \\
      &\leq
      \frac23
      ( \v{b} \otimes \v{z} \otimes \v{z}
      + \v{z} \otimes \v{z} \otimes \v{b}
      + \v{z} \otimes \v{b} \otimes \v{z} )
      && \text{by \cref{eq:mv:AkPosZ2}.}
    \end{align*}
    Moreover, 
    \begin{align*}
      \vv\Lambda^{(3)}\v{x}_3
      &=
      (\vv\Lambda \v{z} \otimes \v{z} \otimes \v{z}
      + \v{z} \otimes \vv\Lambda\v{z} \otimes \v{z}
      + \v{z} \otimes \v{z} \otimes \vv\Lambda \v{z})
      \\
      &>
      (\v{b} \otimes \v{z} \otimes \v{z}
      + \v{z} \otimes \v{b} \otimes \v{z}
      + \v{z} \otimes \v{z} \otimes \v{b}).
    \end{align*}
    Thus,
    $\vv{A}_{33}\v{x}_3 > \frac13 ( (\v{b} \otimes \v{z} \otimes \v{z}
    + \v{z} \otimes \v{b} \otimes \v{z} + \v{z} \otimes \v{z} \otimes
    \v{b}) \geq 0$ and, in turn, $\vv{A}_{33}$ is an $M$-matrix.
  \item $k=4$. %
    Finally, to prove that $\vv{A}_{44}$ is an $M$-matrix, note that
    \begin{align*}
      \vv\Lambda^{(4)}\v{x}_4
      &=
      \vv\Lambda\v{z} \otimes \v{z} \otimes \v{z} \otimes \v{z}
      + \v{z} \otimes \vv\Lambda\v{z} \otimes \v{z} \otimes \v{z}
      \\
      &\qquad\qquad
      + \v{z} \otimes \v{z} \otimes \vv\Lambda\v{z} \otimes \v{z}
      + \v{z} \otimes \v{z} \otimes \v{z} \otimes \vv\Lambda\v{z}
      \\
      &>
      \v{b} \otimes \v{z} \otimes \v{z} \otimes \v{z}
      + \v{z} \otimes \v{b} \otimes \v{z} \otimes \v{z}
      + \v{z} \otimes \v{z} \otimes \v{b} \otimes \v{z}
      + \v{z} \otimes \v{z} \otimes \v{z} \otimes \v{b}
      \intertext{(by \cref{eq:mv:AkPosZ1})}
      &\geq
      \frac{1}{3\lambda_{\min}}\Big(
      \v{b} \otimes \v{b} \otimes \v{z} \otimes \v{z}
      + \v{b} \otimes \v{z} \otimes \v{b} \otimes \v{z}
      + \v{b} \otimes \v{z} \otimes \v{z} \otimes \v{b}
      \\
      &\qquad
      + \v{b} \otimes \v{b} \otimes \v{z} \otimes \v{z}
      + \v{z} \otimes \v{b} \otimes \v{b} \otimes \v{z}
      + \v{z} \otimes \v{b} \otimes \v{z} \otimes \v{b}
      \\
      &\qquad
      + \v{b} \otimes \v{z} \otimes \v{b} \otimes \v{z}
      + \v{z} \otimes \v{b} \otimes \v{b} \otimes \v{z}
      + \v{z} \otimes \v{z} \otimes \v{b} \otimes \v{b}
      \\
      &\qquad
      + \v{b} \otimes \v{z} \otimes \v{z} \otimes \v{b}
      + \v{z} \otimes \v{b} \otimes \v{z} \otimes \v{b}
      + \v{z} \otimes \v{z} \otimes \v{b} \otimes \v{b}
      \Big)
      \intertext{(by \cref{eq:mv:AkPosZ2})}
      &= 
      \frac{2}{3\lambda_{\min}}\Big(
      \v{b} \otimes \v{b} \otimes \v{z} \otimes \v{z}
      + \v{b} \otimes \v{z} \otimes \v{b} \otimes \v{z}
      + \v{b} \otimes \v{z} \otimes \v{z} \otimes \v{b}
      \\
      &\qquad\qquad\qquad
      + \v{z} \otimes \v{b} \otimes \v{b} \otimes \v{z}
      + \v{z} \otimes \v{b} \otimes \v{z} \otimes \v{b}
      + \v{z} \otimes \v{z} \otimes \v{b} \otimes \v{b}
      \Big).
    \end{align*}
    That is,
    $\vv\Lambda^{(4)}\v{x}_4 
    > \frac2{3\lambda_{\min}}\vv{B}^{(4)}(\v{z}\otimes \v{z})$,
    and thus
    \begin{align*}
      \vv{A}_{44}\v{x}_4
      &>
      \vv{B}^{(4)}\Big(
      \frac2{3\lambda_{\min}}(\v{z} \otimes \v{z})
      - (\eye_{p^2}\otimes \v\gamma^T)(\v{z}^{\otimes 4})
      \Big)
      \\
      &= 
      \vv{B}^{(4)}\Big(
      \frac2{3\lambda_{\min}}(\v{z} \otimes \v{z})
      - \v\gamma^T(\v{z}\otimes \v{z})(\v{z} \otimes  \v{z})
      \Big)
      \\
      &= 
      \vv{B}^{(4)}(\v{z}\otimes \v{z})\Big(
      \frac2{3\lambda_{\min}}
      - \v\gamma^T(\v{z}\otimes \v{z})
      \Big)
      \geq 0.
    \end{align*}
    Then, since $\v{x}_4>0$ and $\vv{A}_{44}\v{x}_4 > 0$,
    $\vv{A}_{44}$ is an $M$-matrix.
  \end{itemize}
\end{proof}

\end{document}

%% file: qhr_table.tex
\begin{tabular}{ccrrrrrrrr}
\toprule
 & \multicolumn{4}{l}{Parameters} & \multicolumn{5}{l}{Diagnostics} \\
 & $\lambda$ & $\alpha$ & $\beta$ & $\gamma$ & $2\lambda - \gamma$ & $\sigma_{\min}$ & $\mathbf{y}_{\min}$ & $\sqrt{v_{\infty}}$ & Kurt$_{\infty}$ \\
Model &  &  &  &  &  &  &  &  &  \\
\midrule
$M_1$ & 6.00 & 0.0064 & 0.0000 & 3.6334 & 8.37 & 8.00\% & 0.00 & 9.58\% & 3.00 \\
$M_2$ & 4.00 & 0.0064 & 0.0000 & 2.0000 & 6.00 & 8.00\% & 0.00 & 9.24\% & 1.50 \\
$M_3$ & 6.00 & 0.0133 & -0.1800 & 3.0000 & 9.00 & 5.00\% & 0.06 & 13.32\% & 5.15 \\
$M_4$ & 6.00 & 0.0162 & -0.2280 & 3.8000 & 8.20 & 5.00\% & 0.06 & 15.39\% & 32.29 \\
\bottomrule
\end{tabular}

%% file: r1qhr_table.tex
\begin{tabular}{ccccrrr}
\toprule
 & $\mathbf\Lambda$ & $\mathbf{b}$ & $\mathbf{w}$ & $\alpha$ & $\beta_0$ & $\gamma_0$ \\
Model &  &  &  &  &  &  \\
\midrule
$MM_1$ & $\smx{1 & 0 \\ -1 & 6}$ & $\smx{1 \\ 0}$ & $\smx{0.2 \\ 0.8}$ & 0.0100 & 0.0000 & 2.00 \\
$MM_2$ & $\smx{1 & 0 \\ -1 & 6}$ & $\smx{1 \\ 0}$ & $\smx{0.2 \\ 0.8}$ & 0.0100 & 0.0000 & 2.80 \\
$MM_3$ & $\smx{1 & 0 \\ -1 & 6}$ & $\smx{1 \\ 0}$ & $\smx{0.2 \\ 0.8}$ & 0.0144 & -0.1825 & 2.80 \\
$MM_4$ & $\smx{1 & 0 \\ -1 & 12}$ & $\smx{1 \\ 0}$ & $\smx{0.2 \\ 0.8}$ & 0.0144 & -0.2365 & 4.70 \\
$MM_5$ & $\smx{6 & 0 \\ -1 & 6}$ & $\smx{1 \\ 0}$ & $\smx{1.0 \\ 0.2}$ & 0.0144 & -0.1889 & 3.00 \\
\bottomrule
\end{tabular}

\medskip

\begin{tabular}{crrrrrcrrr}
\toprule
 & $\mu_2$ & $\mu_3$ & $\mu_4$ & $\kappa$ & $\tilde\kappa$ & $\mathbf{y}_{\min}$ & $\sigma_{\min}$ & $\sqrt{v_{\infty}}$ & Kurt$_{\infty}$ \\
Model &  &  &  &  &  &  &  &  &  \\
\midrule
$MM_1$ & 1.75 & 2.28 & 2.60 & 0.10 & 0.49 & $\smx{0.0000 \\ 0.0000}$ & 10.00\% & 10.55\% & 1.03 \\
$MM_2$ & 1.66 & 2.01 & 2.11 & 0.14 & 0.68 & $\smx{0.0000 \\ 0.0000}$ & 10.00\% & 10.79\% & 1.07 \\
$MM_3$ & 1.66 & 2.01 & 2.11 & 0.14 & 0.68 & $\smx{0.0192 \\ 0.0767}$ & 5.00\% & 12.95\% & 1.90 \\
$MM_4$ & 1.66 & 1.98 & 1.99 & 0.16 & 2.26 & $\smx{0.0148 \\ 0.0592}$ & 5.00\% & 13.10\% & 2.17 \\
$MM_5$ & 8.62 & 8.42 & 5.22 & 0.26 & 0.54 & $\smx{0.0606 \\ 0.0121}$ & 5.00\% & 13.94\% & 5.93 \\
\bottomrule
\end{tabular}